\documentclass[manuscript]{emulateapj}

\usepackage[utf8]{inputenc}
\usepackage{graphicx}% Include Figure files
\usepackage{dcolumn}% Align table columns on decimal point

\usepackage{bm}% bold math
\usepackage{xcolor}
\usepackage{color}
\usepackage{verbatim}
\usepackage{lineno}
\usepackage{ulem}
\usepackage{amsmath}
\usepackage[hang]{footmisc}

\usepackage{floatrow}
\usepackage[title]{appendix}

\def\sigint{$\sigma_{\rm int}$}
\def\NSNEPTWENTYONESIMDATA{$1758$}

\def\SIGMAPTWENTYONESIMDATA{$0.0000$}
\def\ALPHAPTWENTYONESIMDATA{$0.140\left(0.003\right)$}
\def\BETAPTWENTYONESIMDATA{$2.774\left(0.037\right)$}
\def\GAMMAPTWENTYONESIMDATA{$0.003\left(0.008\right)$}

\def\RMSPTWENTYONESIMDATA{$0.249$}

\def\NSNEPTWENTYONEDATA{$1792$}

\def\SIGMAPTWENTYONEDATA{$0.0661$}
\def\ALPHAPTWENTYONEDATA{$0.149\left(0.004\right)$}
\def\BETAPTWENTYONEDATA{$3.000\left(0.044\right)$}
\def\GAMMAPTWENTYONEDATA{$0.024\left(0.008\right)$}

\def\RMSPTWENTYONEDATA{$0.285$}

\def\FRACSYSALL{0.7063} 
\def\DELTAWALL{-0.0567}

\def\SIGSYSZEFF{0.005} 
\def\FRACSYSZEFF{0.1243} 
\def\DELTAWZEFF{-0.0036}

\def\SIGSYSMWEBV{0.0025} 
\def\FRACSYSMWEBV{0.0622} 
\def\DELTAWMWEBV{-0.0011} 
\def\SIGSYSCALSALT{0.0185} 
\def\FRACSYSCALSALT{0.4601} 
\def\DELTAWCALSALT{-0.0385} 
\def\SIGSYSCALSPEC{0.0009} 
\def\FRACSYSCALSPEC{0.0224} 
\def\DELTAWCALSPEC{0.0} 
\def\SIGSYSZSHIFT{0.0009} 
\def\FRACSYSZSHIFT{0.0224} 
\def\DELTAWZSHIFT{0.0002} 
\def\SIGSYSMASSLOC{0.0013} 
\def\FRACSYSMASSLOC{0.0323} 
\def\DELTAWMASSLOC{0.0013} 
\def\SIGSYSALPHAEVOL{0.0009} 
\def\FRACSYSALPHAEVOL{0.0224} 
\def\DELTAWALPHAEVOL{-0.0001} 
\def\SIGSYSBETAEVOL{0.0068} 
\def\FRACSYSBETAEVOL{0.1691} 
\def\DELTAWBETAEVOL{0.0012} 
\def\SIGSYSSNN{0.0} 
\def\FRACSYSSNN{0.0} 
\def\DELTAWSNN{0.0} 
\def\SIGSYSCOLORLAW{0.0059} 
\def\FRACSYSCOLORLAW{0.1467} 
\def\DELTAWCOLORLAW{-0.0025} 
\def\SIGSYSVPEC{0.0} 
\def\FRACSYSVPEC{0.0} 
\def\DELTAWVPEC{-0.0001} 
\def\SIGSYSFIXED{0.0054} 
\def\FRACSYSFIXED{0.1343} 
\def\DELTAWFIXED{-0.0047} 
\def\SIGSYSBCONE{0.0055} 
\def\FRACSYSBCONE{0.1368} 
\def\DELTAWBCONE{0.0022} 
\def\SIGSYSBCTWO{0.0059} 
\def\FRACSYSBCTWO{0.1467} 
\def\DELTAWBCTWO{0.0031} 
\def\SIGSYSBCTHREE{0.0089} 
\def\FRACSYSBCTHREE{0.2213} 
\def\DELTAWBCTHREE{0.0017} 
\def\SIGSYS{0.0284}

\def\NSNe{1792}
\def\zmin{0.06}
\def\zmax{0.68}
\def\omegamsn{$0.328 \pm 0.024$}
\def\omegam{$0.330 \pm 0.018$}
\def\w{$-1.016^{+0.055}_{-0.058}$}

\begin{document}
\title{Amalgame: Cosmological Constraints from the First Combined  Photometric Supernova Sample}

\author{Brodie Popovic\footnotemark[1]}
\author{Daniel Scolnic\footnotemark[1]}
\author{Maria Vincenzi\footnotemark[1]}
\author{Mark Sullivan\footnotemark[2]}
\author{Dillon Brout\footnotemark[3]}
\author{Bruno O. Sanchez\footnotemark[1]}
\author{Rebecca Chen\footnotemark[1]}
\author{Utsav Patel\footnotemark[1]}
\author{Erik R. Peterson\footnotemark[1]}
\author{Richard Kessler\footnotemark[4,5]}
\author{Lisa Kelsey\footnotemark[6]}
\author{Ava Claire Bailey\footnotemark[1]}
\author{Phil Wiseman\footnotemark[2]}
\author{Marcus Toy\footnotemark[2]}

\affiliation{$^1$ Department of Physics, Duke University, Durham, NC, 27708, USA.}
\affiliation{$^2$ School of Physics and Astronomy, University of Southampton, Southampton, SO17 1BJ, UK}
\affiliation{$^3$ Department of Astronomy, Boston University, 725 Commonwealth Ave., Boston, MA 02215, USA}
\affiliation{$^5$ Department of Astronomy and Astrophysics, The University of Chicago, Chicago, IL 60637, USA.}
\affiliation{$^6$ Kavli Institute for Cosmological Physics, University of Chicago, Chicago, IL 60637, USA.}
\affiliation{$^7$ Institute of Cosmology and Gravitation, University of Portsmouth, Portsmouth, PO1 3FX, UK}

%\date{}

\begin{abstract}
Future constraints of cosmological parameters from Type Ia supernovae (SNe Ia) will depend on the use of photometric samples, those samples without spectroscopic measurements of the SNe Ia. There is a growing number of analyses that show that photometric samples can be utilised for precision cosmological studies with minimal systematic uncertainties. To investigate this claim, we perform the first analysis that combines two separate photometric samples, SDSS and  Pan-STARRS, without including a low-redshift anchor. We evaluate the consistency of the cosmological parameters from these two samples and find they are consistent with each other to under $1\sigma$. From the combined sample, named Amalgame, we measure $\Omega_M = $ \omegamsn\ with SN alone in a flat $\Lambda$CDM model, and $\Omega_M = $ \omegam\ and $w =$ \w\ when combining with a Planck data prior and a flat $w$CDM model. These results are consistent with constraints from the Pantheon+ analysis of only spectroscopically confirmed SNe Ia, and show that there are no significant impediments to analyses of purely photometric samples of SNe Ia. 
\end{abstract}

\maketitle

\section{Introduction}
Type Ia supernovae (SNe Ia) are a crucial component of cosmological analyses due to their unparalleled mapping of the universe's expansion history. The accelerating expansion of the universe was discovered with the use of SNe Ia by \cite{Riess98} and \cite{Perlmutter99}, but the cause of this acceleration remains an unsolved mystery in cosmology. In the intervening decades, statistical and systematic improvements in SNIa samples have increased the precision of measurements of the Dark Energy equation-of-state $w=(P/\rho c^2)$; most recently \cite{Brout22} find $w=-0.978^{+0.024}_{-0.031}$ when combined with constraints from the Cosmic Microwave Background radiation \citep{Planck18} and Baryon Acoustic Oscillations \citep{Ross15, Alam17, Bautista20, Hou20, Chabanier21}. 

The strength of SNe Ia come from their status as standardisable candles - their luminosity, which is enough to still be seen at cosmic distances, can be standardised to $\sim 0.1 $ magnitudes, which corresponds to $\sim$5\% in distance per supernova. Statistical improvements in SNIa constraints have come in two parts. The first is the addition of new samples, such as the Sloan Digital Sky Survey (SDSS; \citealp{Sako11, Campbell13}), the Supernova Legacy Survey (SNLS; \citealp{Astier06}), Pan-STARRS (PS1; \citealp{Scolnic18}), the Dark Energy Survey (DES; \citealp{DES3YR}), and Foundation \citep{Foley18}. 
The second of these statistical improvements comes from the compilation of these surveys such as Union \citep{Union}, the Joint Light-curve Analysis \citep{Betoule14}, the Pantheon \citep{Scolnic18} and Pantheon+ \citep{Brout22} samples, the latter of which is the largest to-date supernova collection of 1550 unique spectroscopically-confirmed SNIa. The practice of combining samples for improved statistical constraints has become commonplace in recent years.

However, these samples have relied on spectroscopically confirmed SNe Ia and {to date, a compilation of multiple photometric samples has not been done}. While the practice of spectroscopically confirming SNIa has the benefit of reducing contamination from other types of SNe (which are not standardisable in the same way as SNIa) and confirming the redshift of the supernova, spectroscopic confirmation for recent surveys has only been available for $\sim10\%$ of observed SNIa (e.g. SDSS, \citealp{Popovic19}). 

Photometrically classified samples are typically much larger, but cosmologists must still contend with non-Ia contamination. To mitigate this problem, photometric classifiers such as Nearest Neighbor \citep{Sako18, Kessler16}, PSNID \citep{Sako08}, SuperNNova \citep{Moller19}, and SCONE \citep{Qu21} are used to assign a probability for multiple SN types. These classifiers can be used in conjunction with the Bayesian Estimation Applied to Multiple Species (BEAMS, \citealp{Hlozek12}) method to rigorously account for non-SNIa contamination by weighting each distance measurement by their likelihood of being type Ia. The BEAMS method has been implemented and tested in several analyses \citep{Kunz07, Jones18, Jones19, Vincenzi21} and has been proved to reproduce unbiased cosmological results.

Here we focus on two photometric samples that have have been used in previous SNIa-cosmology analyses: the PS1 photometric sample \citep{Jones18} and the SDSS photometric sample (\citealp{Sako16} with updates from \citealp{Popovic19}). These samples contain respectively $\sim 1200$ and $\sim 700$ likely SNIa that pass light curve fitting, larger than any individual spectroscopic survey. We compare Hubble Residuals (the difference between SNIa distance and the best-fit cosmology), light-curve fitting parameters, and measure the dark energy equation-of-state parameter $w$. These two subsamples comprise the Amalgame sample; in contrast to the historical SNIa analyses listed above, Amalgame does not include spectroscopically confirmed SNe Ia at low redshift $(z < 0.1)$

This paper will set the groundwork for combining newer photometric samples and pave the way for a combined (e.g. PS1, SDSS, and DES) photometric sample that can comprise $\sim 4,000$ photometrically classified type Ia supernovae. The layout of the paper is as follows. Section \ref{sec:SNInfo} details methodologies such as distance measurements, bias corrections, non-Ia contamination, Bayesian Estimation Applied to Multiple Species, covariance matrices,  cosmological calculations, and simulations. Section \ref{sec:Inputs} contains an overview of the data and systematics, and Section \ref{sec:Results} contains the cosmological results. Finally, Section \ref{sec:Discussion} includes the discussion and Section \ref{sec:Conclusion} details our conclusions.

\section{Constraining Cosmological parameters with SNIa}\label{sec:SNInfo}

\subsection{Data}\label{sec:SNInfo:subsec:data}

The PS1 survey operated from 2009 to 2013, with a cadence of roughly 6 observations every 10 days in 10 7-square degree fields with 5 filters. Light curves for PS1 are taken from \cite{Jones18}.\footnote{https://archive.stsci.edu/prepds/ps1cosmo/} The PS1 data were taken with the \textit{griz}\footnote{\textit{Y} band data was taken, but is not used here.} filters \citep{Chambers16}, host galaxy redshifts are drawn from a compilation of sources detailed in \cite{Jones18}.

The SDSS supernova program ran for three observing seasons from 2005 to 2007 with a cadence of observing every four days with five filters. An overview of the SDSS survey is available in \cite{Frieman08} and the \textit{ugriz} filters are detailed in \cite{Doi10}. We do not include the $u$-band data, due to issues with standardisation detailed in \cite{fragilistic}. The SDSS light curves are taken from \cite{Sako18}, though the host galaxies and their associated redshifts were updated in \cite{Popovic19}.

Host galaxy stellar masses for SDSS and PS1 are rederived for this paper from their photometry releases following the approach detailed in \cite{Smith20}. Full details can be found in Appendix \ref{sec:HostMass}.

While both SDSS and PS1 used similar image-subtraction techniques to discover SNe, they used different methods to provide accurate flux measurements for light curves used in their cosmology analyses. \cite{Holtzman08} used Scene Modeling Photometry (SMP) for the SDSS sample, whereas \cite{Jones18} used difference imaging. \cite{Scolnic18} compared the impact of these different image subtraction techniques and found them to be consistent.

\subsection{Distances}\label{sec:SNInfo:subsec:distance}

SNIa brightnesses are standardised with the use of a light-curve model and fitting process; here we use the SALT3 model as developed by \cite{Kenworthy21} based on the SALT2 model from \cite{Guy10}. The SALT3 fit reports four parameters for each supernova: An overall light-curve amplitude $x_0$, a colour parameter $c$, a stretch parameter $x_1$ related to the light-curve width, and the time of peak brightness $t_0$. The stretch and colour parameters are used to standardise the SNIa brightness, and the distance $\mu$ is determined from a modified version of the Tripp estimator \citep{Tripp98, Scolnic18}:

\begin{equation}\label{eq:tripp}
\mu = m_B + \alpha x_1 - \beta c - M_0 - \delta \mu_{\rm host} - \delta \mu_{\rm bias} 
\end{equation}
where $m_B = -2.5\log_{10}(x_0)$, $c$ and $x_1$ are defined above, and $M_0$ is the absolute magnitude of a SNIa with $c = x_1 = 0$. $\delta \mu_{\rm host}$ accounts for residual standardised brightness dependencies between the SNIa and its host galaxy; here the host-galaxy stellar mass:
\begin{equation}
    \delta \mu_{\rm host} = 
    \begin{cases} 
      ~+\gamma/2  & \textrm{if}~ M_{*} > M_{\rm step}\\
      ~-\gamma/2  & \textrm{if}~ M_{*} \leq M_{\rm step}
   \end{cases}
\label{eq:gamma}
\end{equation}
where $\gamma$ is the magnitude difference, typically $\sim 0.05$ magnitudes \citep{Sullivan10}, $M_{*}$ is the host-galaxy stellar mass, and $M_{\rm step}$ is the location of the `mass step'. $\alpha$ and $\beta$ are the stretch-luminosity and colour-luminosity nuisance parameters. Finally, $\delta\mu_{\rm bias}$ is a bias correction determined from simulations. 

\begin{table}[h]
    \centering
    \begin{tabular}{c|c|c}
         Survey & Before Cuts & After Cuts \\
         \hline
         SDSS & 2015 & 646 \\
         PS1 & 2478 & 1153 \\
         Total & 4493 & 1792
    \end{tabular}
    \caption{Number of SNe that pass SALT3 fitting before and after cosmology cuts.}
    \label{tab:numbers}
\end{table}

The cosmology cuts \citep{Betoule14} instituted for our data and simulations are shown below; Table \ref{tab:numbers} shows a breakdown of the number of likely SNIa that pass through light curve fitting before and after our cosmology cuts. 

\begin{itemize}
    \item $\sigma_{x_1} < 1$ : SALT3 $x_1$ uncertainty $< 1$.
    \item $\sigma_{\rm PKMJD} < 2$ : Uncertainty on fitted peak brightness $< 2$ days.
    \item $-3< x_1 < 3$.
    \item $-0.3 < c < 0.3$.
    \item $T_{\rm rest, min} < 5$ : Requires at least 1 observation 5 days before peak brightness (rest frame).
    \item $T_{\rm rest, max} > 0$ : Requires at least 1 observation after peak brightness (rest frame).
    \item $ -20 < T_{\rm rest} < 60$ : Requires at least one observation between -20 and 60 days (rest frame).
\end{itemize}

SDSS, in particular, loses over 50\% of the potential SNe Ia after cosmology cuts; these losses are primarily due to $x_1$ values outside of the range of $-3 < x_1 < 3$ and error cuts such as $\sigma_{\rm PKMJD}$ and $\sigma_{x_1}$; e.g., noisier data in SDSS.

\subsection{BEAMS with Bias Corrections}\label{sec:SNInfo:subsec:bbc}
SNIa distance measurements can be biased due to Malmquist bias and analysis selection effects \citep{Kessler09, Betoule14, Scolnic16, Popovic21a}. In this analysis, $\delta \mu_{\rm bias}$ is determined from simulations using the BEAMS with Bias Corrections (BBC) framework from \cite{Kessler16} with updates from \cite{Popovic21a}. BBC is designed to correct distance biases arising from selection effects and to account for the presence of non-Ia supernovae in a cosmological analysis; the component that handles non-Ia contamination is the Bayesian Estimation Applied to Multiple Species (BEAMS) framework, and is detailed further in Section \ref{sec:SNInfo:subsec:cc}. The bias-corrections aspect of BBC works by including distance corrections determined from the use of rigorous simulations. BBC fits for nuisance parameters $\alpha$, $\beta$, $\gamma$, and $\sigma_{\rm int}$ as described in \cite{Marriner11}, and produces a binned and unbinned Hubble diagram \citep{Kessler23, Kessler16} that is corrected for selection effects and non-Ia contamination.

The BBC-4D framework (BBC-BS20, \citealp{Popovic21a}) models $\delta \mu_{\rm bias} = \delta \mu_{\rm bias} \{z, c, x_1, \log M_{*}\}$ (where $M_{*} = M_{\rm stellar}/M_{\rm sun}$) to account for dust-based models of SNIa scatter. The measured distances in dust-based models are affected by the dust law of the SN host galaxy, an effect seen in \cite{BS20}, Pantheon+ (\citealp{Popovic22, Brout22}) and DES (\citealp{Wiseman22, Kelsey22}). The SALT light curve fitter does not directly measure nor infer the dust properties themselves (e.g. $R_V$, $A_V$, $E(B-V)$), so BBC-4D is not directly able to correct for these, but rather makes use of simulations (Section \ref{sec:SNInfo:subsec:simulations}) to correct for observed SNIa properties. This paper includes updates to the BBC-4D framework, detailed in Appendix \ref{appendix:sigint}.

\subsubsection{Simulations}\label{sec:SNInfo:subsec:simulations}

To compute the bias correction term $\delta\mu_{\rm bias}$, simulations of supernovae are necessary. The simulation generates catalogue-level SNIa light curves in three general steps:

\begin{itemize}
    \item 1. True broadband fluxes are generated from a model. 
    \item 2. Noise is applied.
    \item 3. Detections are determined based on telescope and survey-specific criteria. 
\end{itemize}

Firstly, a rest-frame Spectral Energy Distribution (SED) is created for each SNIa epoch and cosmological effects (redshift, dimming) and galactic effects (Galactic extinction, peculiar velocities, lensing) are applied. An additional model is needed to accurately capture the scatter in SNIa distances that is not due to measurement error; this is included here. This rest-frame SED is integrated for each telescope filter to obtain broadband photometry fluxes. 

It is at this step that the intrinsic distributions of SNIa, such as colour and stretch, are selected from a population given in the simulation input. The measurements of these intrinsic distributions result in skewed and smeared populations affected by measurement noise and detection efficiency (steps two and three).

Measurement error is applied from the source flux, zero points, PSF, and sky noise. Thirdly, detections are determined through the application of survey-dependent estimate of efficiency vs. signal-to-noise. For photometric samples, the simulation includes an additional efficiency for measuring an accurate spectroscopic host-galaxy redshift. This is typically a function of the host galaxy brightness (\citealp{Popovic19, Vincenzi21}).

%%%%%%%%%%%%%%%%%%%%%%

\begin{table*}[!t]
    \centering
    \begin{tabular}{c|c|c|c}
        Survey & Cadence & Detection Eff. & Spectroscopic Eff. \\
        \hline
        SDSS & \cite{Sako18, Kessler13} & This work & This work \\ %\cite{Sako18, SNANA}
        PS1  & \cite{Jones18} & \cite{Jones18} & This work \\
    \end{tabular}
    \caption{Review of Simulation Inputs for Amalgame}
    \label{tab:siminputs}
\end{table*}

\begin{figure}[h]
\includegraphics[width=8cm]{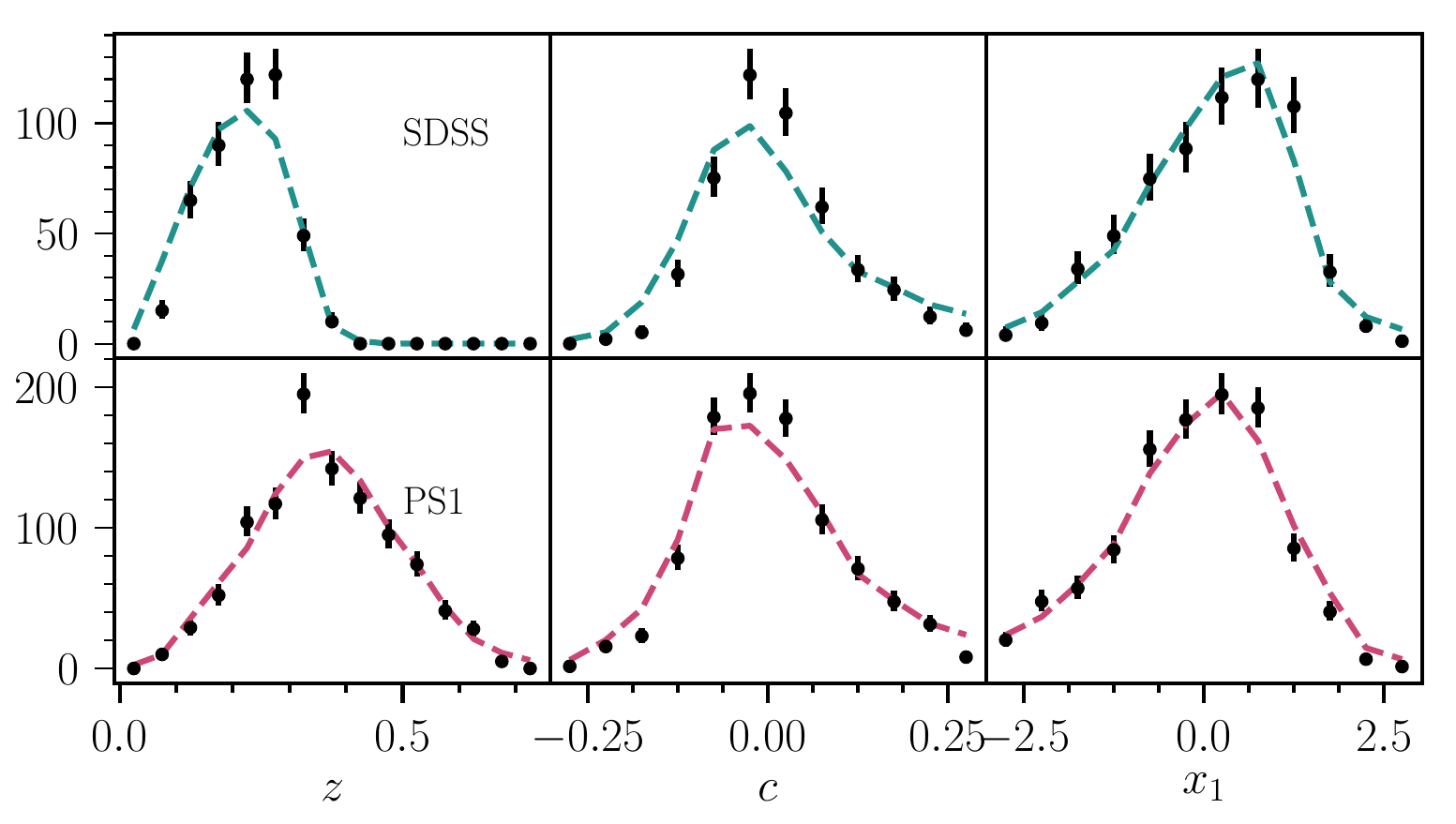}
\caption{Comparison between simulations in dashed histogram and data in black points for the subsamples in this analysis: SDSS and PS1. Three distributions are compared: $z$, $c$, and $x_1$. More simulation-data comparison plots are found in the Appendix.}
\label{fig:basic_hists}
\end{figure}

SDSS and PS1 are modeled for the simulations with the use of survey metadata -- logs of observations, including sky noise, zero points, and other telemetry are included within the simulation. Table \ref{tab:siminputs} shows the origin of these simulation inputs. Figure \ref{fig:basic_hists} shows good agreement between the data and our overlaid simulations for SDSS and PS1 for redshift, $c$, and $x_1$.

\subsubsection{SDSS Detection Efficiency}\label{sec:Discussion:DETEFF}

\begin{figure}[h]
\includegraphics[width=8cm]{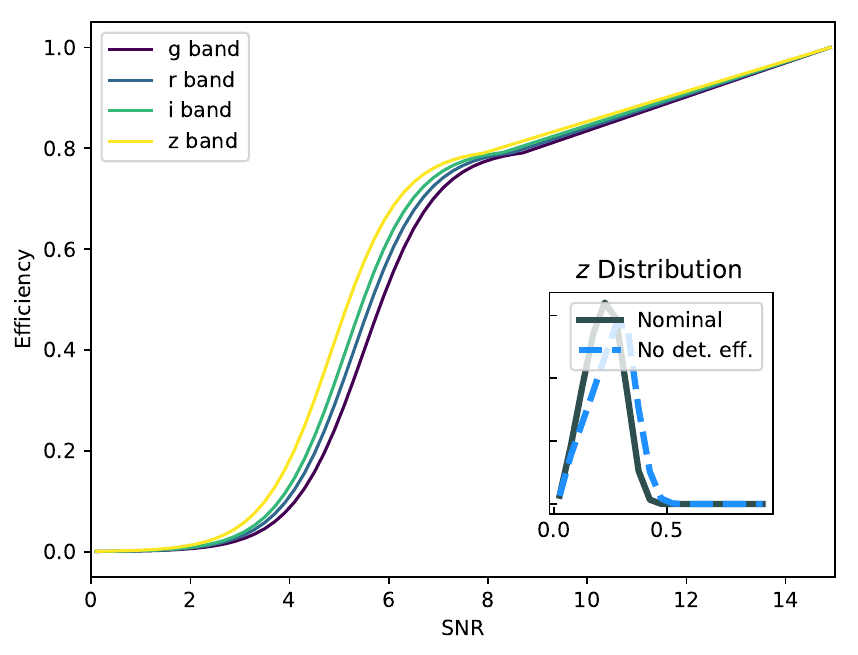}
\caption{The SDSS detection efficiency for simulated supernovae as a function of SNR. The g band is presented in purple, r band in blue, i band in green, and z band in yellow. In the inset plot is the redshift distribution for the nominal simulations (grey histogram) and no detection efficiency simulation (blue dashed histogram).}
\label{fig:DETEFF}
\end{figure}

In the course of the Amalgame analysis, a new detection efficiency was created using the results from \cite{Sanchez21}. This detection efficiency, shown in Figure \ref{fig:DETEFF}, was used in the fiducial sample, and a systematic with increased efficiency was included to approximate the impact on measuring cosmological parameters.  

%%%%%%%%%%%%%%%%%%%%%%%%%%%%%%%%%%%%%%%%%%%%%%%%%%%%%%%%%%%%%%%%%%%%
%%%%%%%%%%%%%%%%%%%%%%%%%%%%%%%%%%%%%%%%%%%%%%%%%%%%%%%%%%%%%%%%%%%%

\subsubsection{Error Modeling}\label{sec:SNInfo:subsec:error}

After bias corrections are calculated from simulations, BBC determines the distance uncertainties and calculates $\sigma_{\rm int}$, the remaining post-standardisation scatter that cannot be attributed to known experimental sources of noise.
The error on the distance modulus, $\sigma_\mu$, is 
\begin{multline}
        \sigma_{\mu}^2 = f(z,c,M_{\star})\sigma^2_{{\rm meas}} + \sigma^2_{\rm int}(z,c,M_{\star}) \\ + \sigma^2_{{\rm lens}} + \sigma^2_{{\rm vpec} \, ,} 
    \label{eq:tripperr}
\end{multline}
where the measurement uncertainty of light-curve fit parameters and their covariances from SALT3 is represented by $\sigma_{\rm meas}$. This measurement uncertainty is scaled by a survey-specific amount $f(z,c,M_{\star})$ to account for selection effects that suppress distances corresponding to fainter magnitudes. $\sigma_{\textrm{lens}} = 0.055z$ -- the effect of gravitational lensing on SNIa brightness -- is taken from \cite{Jonsson10}.  $\sigma_{\rm vpec}$ is the distance uncertainty arising from peculiar velocities (set to $250~\textrm{kms}^{-1}$ as per Pantheon+). 

As stated above, \sigint\ is the remaining scatter after standardisation, and is modeled as 
\begin{equation}
\label{eq:errmodel}
   \sigma^2_{\rm int}(z,c,M_{\star}) = \sigma^2_{\rm scat}(z,c,M_{\star}) + \sigma^2_{\rm gray},
\end{equation}
where $\sigma^2_{\rm scat}(z,c,M_{\star})$ is a SN-property dependent value and $\sigma^2_{\rm gray}$ is a scatter floor for all SNe~Ia. The $\sigma^2_{\rm gray}$ is determined to set the reduced $\chi^2$ from the \cite{Marriner11} fit to 1. Further details on $\sigma^2_{\rm scat}(z,c,M_{\star})$ can be found in Section 3.3 and Appendix A of \cite{Brout22}.

%%%%%%%%%%%%%%%%%%%%%%%%%%%%%%%%%%%%%%%%%%%%%%%%%%%%%%%%%%%%%%%%%%%%
%%%%%%%%%%%%%%%%%%%%%%%%%%%%%%%%%%%%%%%%%%%%%%%%%%%%%%%%%%%%%%%%%%%%

\subsection{Core Collapse Supernova}\label{sec:SNInfo:subsec:cc}

After light curve fits and conventional quality cuts, some non-Ia Sne may appear to be SNIa; approximately 1\% of the resulting sample is likely to be non-Ia contamination \citep{Vincenzi21}. To account for potential biases from non-Ia contamination we make use of the Bayesian Estimation Applied to Multiple Species (BEAMS) framework presented in \cite{Hlozek12} and updated in \cite{Kessler16}.

The BEAMS approach marginalises over non-Ia contamination while performing the cosmological fit. This is done within the BBC framework \citep{Kessler16}. BEAMS uses two likelihoods to model distinct SN populations. The first likelihood models the SNIa population $\mathcal{L}_\mathrm{Ia}$ and the second models the non-Ia contaminants  $\mathcal{L}_\mathrm{CC}$ such that 

\begin{equation}\label{eq:BEAMS1}
    \mathcal{L}_{\rm tot} = \sum_{i=1}^{N_{\mathrm{SNe}}} \mathcal{L}^i_\text{Ia}+\mathcal{L}^i_\text{CC}.
\end{equation}

These likelihoods are defined as 

\begin{equation}
  \begin{aligned}
    \mathcal{L}^i_\text{Ia} &=  \tilde{P}^i_{\text{Ia}} \times \text{exp} \biggl(-\frac{(\mu_{\text{obs},i} + \Delta \mu^{i} - \mu_{\text{ref}}(z_i))^2} {\sigma_{\mu,i}^{2}} \biggr)\\
    \mathcal{L}^i_\text{CC} &= (1-\tilde{P}^i_{\text{Ia}}) \times D_{\text{CC}}(z_i, \mu_{\text{obs},i}, \mu_{\text{ref}}),
  \end{aligned}
  \label{eq:likelihoods}
\end{equation}

where the distance modulus of a given SN $i$ is predicted assuming a fixed reference cosmology -- an arbitrary placeholder cosmology that will be marginalised out during the full cosmological fit -- and $\Delta \mu$ is the offset from the reference cosmology. $\tilde{P}^i_{\text{Ia}}$ is the scaled likelihood of the $i$th SN being a type Ia supernova, and $D_{\text{CC}}$ is the contamination likelihood term, given in Equation 5 of \cite{Vincenzi21}. The likelihood in Equation \ref{eq:BEAMS1} is maximised, and the distance modulus uncertainties, $\sigma_{\mu, i}$, are determined in conjunction with Equation \ref{eq:tripperr}. For those SNe with lower $P_{Ia}$ values, the distance uncertainty is increased via the BBC process during the cosmological fit to account for non-Ia contamination \citep{Kessler23}.

\subsection{Covariance Matrix}\label{sec:SNInfo:subsec:covariance}

%%%%%%%%%% SYSTEMATICS %%%%%%%%%%%%%%%%%%%%%  

We compute statistical and systematic covariance matrices, following \cite{Conley11}, to account for the correlations between SNIa light curves and the systematic and statistical uncertainties in our analysis. We use the unbinned covariance matrix approach from \cite{Binning} with uncertainty scales from \cite{Kessler23} to avoid inflating systematics with a redshift-binned Hubble diagram. 
We define our statistical covariance matrix $C_{\rm stat}$, taking into account that the same SN can be observed in two different surveys, as
\begin{equation}
        C_{\rm stat}(i,j) =
        \begin{cases}
            \vspace{3mm}
            \sigma_\mu^2 & i=j \\
            \sigma^2_{\rm floor}+ \sigma^2_{\rm lens}+\\ \sigma^2_z+\sigma^2_{\rm vpec} & i\neq j~\&~{\rm SN}_i={\rm SN}_j
        \end{cases} ,
\end{equation}
where each row of $C_{\rm stat}$ corresponds to an SN light curve, $i$ and $j$ are the column and row indices for each supernova, and the diagonal is the full distance error. 
The process of taking SNIa light curves to cosmological measurements invites a number of systematic uncertainties. The largest categories of systematic uncertainty are:

\begin{itemize}
    \item Assumptions and calibration related to the light-curve fitting. 
    \item Redshift measurements. 
    \item Astrophysical or survey-dependent factors in the bias correction simulations.
\end{itemize}

Each of these broad categories needs to be accounted for individually to determine the extent of the systematic uncertainty $\psi$ and the size of the systematic uncertainty S$_\psi$. For the covariance matrix, we define the set of distance uncertainties for a given systematic $\psi$:
\begin{equation}
\label{eq:deltamupsi}
\Delta\mu^i_\psi \equiv \mu^i_\psi - \mu^i_{\rm BASE} - ( \mu_{\rm ref}(z_\psi)-\mu_{\rm ref}(z_{\rm BASE}) ) 
\end{equation} 
where $\mu_{\rm BASE}$ is the SNIa distance that corresponds to using $z_{\rm BASE}$, the base redshift, and $\mu_{\psi}$ is similarly the distance corresponding to using systematic redshift $z_{\psi}$. Systematics that affect redshift, such as changing the host-galaxy followup efficiency, include the methodology from Equation 6 in \cite{Brout22} that propagates the effects of changing redshifts into the covariance matrix by accounting for the difference in model differences ($\mu_{\rm mod}(z_\psi)-\mu_{\rm mod}(z_{\rm BASE})$)).

We compute the distance shift for each $\psi$, and use it to build a \NSNe$\times$\NSNe~systematic covariance matrix for each supernova as 
\begin{equation}
\label{eq:csys}
C^{ij}_{\rm syst} = \frac{1}{W_{\psi}} \sum_{\psi} \Delta\mu^i_\psi \times \Delta\mu^j_\psi
\end{equation}
which gives the covariance between the $i^{\rm th}$ and $j^{\rm th}$ supernova, marginalised over the systematic uncertainties $\psi$ with their uncertainty, $\sigma_{\psi}$. The $\frac{1}{W_{\psi}}$ is included as an optional weight for the systematic, for including multiple realisations of a single systematic (realisations of different dust models (Section \ref{sec:Inputs:subsec:scattermodel}) or SALT surfaces (Section \ref{sec:Inputs:subsec:salt3})). To maintain a consistent sample of SNIa for accurate comparisons of systematics, we constrain our final set of SNe Ia to be the subset of SNe Ia that both pass quality cuts and have valid bias corrections across all the tested systematics. This results in a loss of $\sim 250$ SNe. 

When constraining cosmological parameters, we combine the systematic covariance matrix with the statistical covariance matrix: 
\begin{equation}
\label{eq:cstatplussyst}
C_{{\rm stat+syst}} = C_{\rm stat} + C_{{\rm syst}}.
\end{equation}

\subsection{Cosmology Fitting}\label{sec:SNInfo:subsec:cosmology}

We constrain our cosmological parameters using the $\chi^2$ likelihood method from \cite{Conley11}, minimising 
\begin{equation}
\label{eq:likelihood}
-2{\rm ln}(\mathcal{L}) = \chi^2 = \Delta \vec{D}^T~C_{\rm stat+syst}^{-1}~\Delta \vec{D} ,
\end{equation}
where our \NSNe~ SNIa distance modulus residuals is $\Delta \vec{D}$ with components
\begin{equation}
\label{eq:dmu}\Delta D_i = \mu_i - \mu_{{\rm model}}(z_i) .
\end{equation}
The predicted distance at a given redshift from the cosmological fit ($\mu_{{\rm model}}(z_i)$) is subtracted from the individual SNIa distance measurement $\mu_i$. The model distance is given as a function of the luminosity distance $d_L$,
\begin{equation}
\mu_{{\rm model}}(z_i) = 5\log(d_L(z_i)/10\,{\rm pc}) ,
\end{equation} 
and includes the expansion history $H(z)$.

For the flat cosmologies that we examine ($\Omega_k=0$), this luminosity distance is 

\begin{equation}
\label{eq:dl}
d_L(z) = (1+z)c\int_0^{z}\frac{dz^\prime}{H(z^\prime)},
\end{equation}
and $d_L(z)$ is iteratively calculated at each step of the cosmology fitting process. The expansion history $H(z)$, used in Equation \ref{eq:dl} and subsequently in Equation \ref{eq:likelihood}, is 

\begin{equation}
H(z) = {{\rm H}_0}\ \sqrt[]{\Omega_M(1+z)^3+\Omega_{\Lambda}(1+z)^{3(1+w)}}.
\label{eq:hz}
\end{equation}

We assume two models of cosmology for this analysis:\
\begin{itemize}
    \item $\Lambda$CDM; float $\Omega_M$ and $\Omega_\Lambda$, fix $w = -1$
    \item Flat$w$CDM; float $w$ and $\Omega_M$, fix $\Omega_M + \Omega_\Lambda = 1$
\end{itemize}

We also include a prior from the Cosmic Microwave Background radiation (CMB, \citealp{Planck18}).

\subsection{Software Programs}

This analysis is performed using the SNANA software package \citep{SNANA, Kessler19} with use of the PIPPIN software package \citep{PIPPIN}. PIPPIN includes SNANA integration with the 
the photometric classifier  SuperNNova \citep{Moller19} to assign a probability for each event to be of type SNIa. The SuperNNova classifier trains and classifies directly on SN light curves through the use of a recurrant neural network. 

For our final cosmology contours, we use CosmoSIS \citep{zuntz}. numerical estimates of specific $w$-uncertainties are performed with \texttt{wfit} \citep{Kessler19}. 

%%%%%%%%%%%%%%%%%%%%%%%%%%%%%%%%%%%%%%%%%%%%%%%%%%%%%%%%%%%%%%%%%%
%%%%%%%%%%%%%%%%%%%%%%%%%%%%%%%%%%%%%%%%%%%%%%%%%%%%%%%%%%%%%%%%%%

\section{Systematic Uncertainty Sources} \label{sec:Inputs}

Here we introduce and detail the assumptions and approaches in our cosmological analysis that lead to systematic uncertainties. Each assumption is discussed and motivated, along with a complementary method of modeling to account for systematic uncertainties. The \uline{\textit{purpose}}, \uline{\textit{baseline}} treatment, and \uline{\textit{systematic}} uncertainties are given. The effect of these systematics are shown in Section~\ref{sec:Results}; the detailed description of the systematics themselves are given here.

\textbf{Data} \\
Section \ref{sec:Inputs:subsec:hostgal} Host-Galaxy Properties (Also see \ref{sec:Inputs:subsec:Other}) \\

\textbf{Calibration and Light-Curve Fitting} \\
Section \ref{sec:Inputs:subsec:calibration} Calibration \\
Section \ref{sec:Inputs:subsec:salt3} SALT3 Model \\

\textbf{Simulations} \\
Section \ref{sec:Inputs:subsec:surveymodel} Survey Modeling \\
Section \ref{sec:Inputs:subsec:scattermodel} Intrinsic Scatter Models \\
Section \ref{sec:Inputs:subsec:ccprior} Core Collapse Models \\

\textbf{Other} \\
Section \ref{sec:Inputs:subsec:nuisanceparams} Evolution of Nuisance Parameters with $z$ \\
Section \ref{sec:Inputs:subsec:nuisancefixed} Fixed Nuisance Parameters \\
Section \ref{sec:Inputs:subsec:validation} Validation \\
Section \ref{sec:Inputs:subsec:Other} Other \\ 

%%%%%%%%%%%%%%%%%%%%%%%%%%%%%%%%%%%%%%%%%%%%%%%%%%%%

\begin{figure}[h]
\includegraphics[width=8cm]{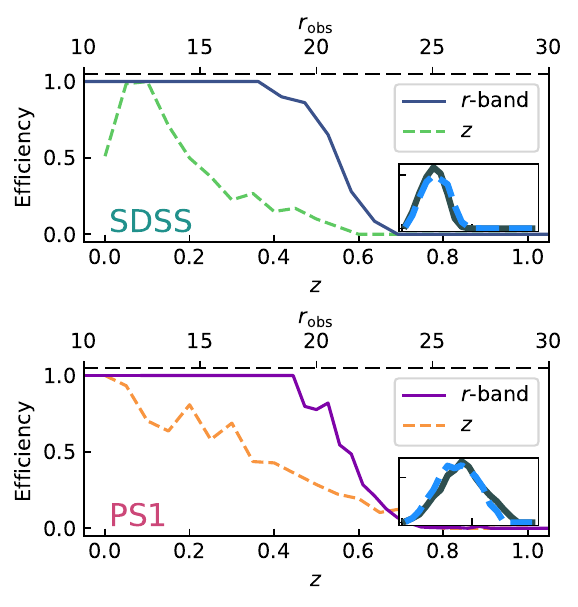}
\caption{A comparison of the host galaxy redshift efficiencies for SDSS (top) and PS1 (bottom). The $r$-band magnitude efficiency used in the fiducial analysis is presented in solid line, and the $z$ efficiency is presented in dashed line. The resulting redshift distributions are presented in the inset plots for the $r$-band magnitude in solid histogram and $z$ efficiency in dashed histogram.}
\label{fig:EFF_COMP}
\end{figure}

\subsection{Host-Galaxy Properties}\label{sec:Inputs:subsec:hostgal}

\uline{\textit{Purpose}}: The host-galaxy mass is used to standardise and improve SNIa distances in two ways: simulations include correlations between $x_1$ and the host-galaxy mass from \cite{Popovic21a} and correlations between $c$, dust, and host-galaxy mass from \cite{Popovic22}; and a residual luminosity step related to the host galaxy mass (the `mass step' $\gamma$) is corrected in the Tripp Eq. \ref{eq:tripp} (Appendix \ref{appendix:Dust2Dust}). Because SALT does not account for covariances between SNIa fitted properties and host galaxies, the first of these use cases accounts for observed correlations between fitted SNIa properties $x_1$ and $c$ in the data. These correlations are incorporated into the simulations and bias corrections so as not to suppress nuisance parameters as in \cite{Smith20}.

\uline{\textit{Baseline}}: We rederive the host-galaxy stellar masses for all host galaxies using the methodology laid out in Appendix \ref{sec:HostMass}. For the baseline analysis, we place the mass step at $10^{10} M_{\odot}$. 

\uline{\textit{Systematics}}: Analyses such as \cite{Sullivan11}, \cite{Childress14}, and \cite{Kelsey20} have found evidence that a more appropriate location of the mass step may not be at $10^{10} M_{\odot}$. Following Pantheon+, we include a systematic placing the mass step at $10^{10.2} M_{\odot}$.

%%%%%%%%%%%%%%%%%%%%%%%%%%%%%%%%%%%%%%%%%%%%%%%%%%%%

\begin{figure}[h]
\includegraphics[width=9cm]{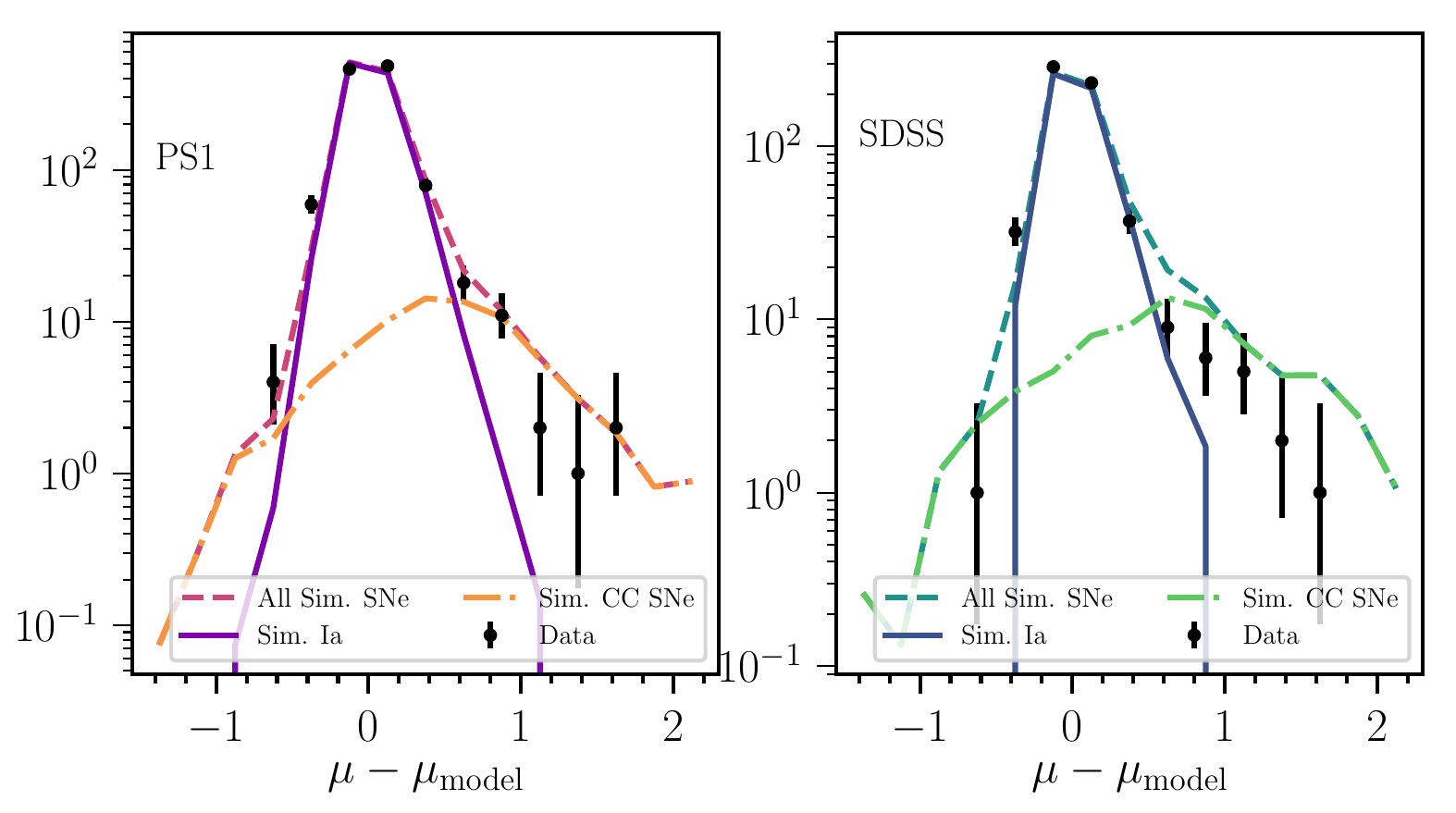}
\caption{Hubble Residuals for PS1 (left) and SDSS (right). The data is shown in black points, simulations are broken down into `All Simulated SNe' in dashed histogram, `Simulated Ia Only' in solid histogram, and only `Simulated Core Collapse Only' in dash-dotted histogram. Both surveys show good agreement between sim and data.}
\label{fig:MURESCC}
\end{figure}

%%%%%%%%%%%%%%%%%%%%%%%%%%%%%%%%%%%%%%%%%%%%%%%%%%%%

\subsection{Calibration}\label{sec:Inputs:subsec:calibration}
\uline{\textit{Purpose:}} The compilation of different instruments, filters, and telescopes requires photometric calibration of the passbands in each survey in order to fit light curves and standardise SN brightnesses. This same calibration is needed for the SALT3 model training.

\uline{\textit{Baseline:}} We use the calibration presented by Fragilistic \citep{fragilistic}. Alongside the nominal calibration, we make use of the relevant subset of their $105 \times 105$ covariance matrix that describes the zeropoint calibrations. This $105 \times 105$ matrix includes the uncertainties and zero points and effective wavelengths of the photometric bands from SDSS and PS1.

\uline{\textit{Systematics:}} A more in-depth description of the calibration and its systematics is given in \cite{fragilistic}. To estimate the calibration systematic, we refit SALT3 for 9 realisations of calibration zero points drawn from the Fragilistic covariance matrix. The value of $W_{\psi}=1/9$ is chosen such that the sum of calibration weights is unity. Additional changes to the wavelength range and tertiary stellar magnitudes are given in Section 3.2.1 in \cite{Brout22}.

%%%%%%%%%%%%%%%%%%%%%%%%%%%%%%%%%%%%%%%%%%%%%%%%%%%%

\subsection{SALT3 Model} \label{sec:Inputs:subsec:salt3}
\uline{\textit{Purpose}}: SNIa light curves are fit with the SALT3 model to determine the light-curve parameters $m_B, c, x_1$ for each SN for use in Equation \ref{eq:tripp}.

\uline{\textit{Baseline}}: We use the SALT3 model from \cite{Kenworthy21} with the calibration from \cite{fragilistic}. We use a SALT3 surface trained without \textit{U}-band, after finding potential cosmological biases due to mis-calibrated \textit{U}-band information. This is further detailed in \cite{DES5YR}.

\uline{\textit{Systematics}}: Following \cite{Brout22}, we use 9 SALT3 models retrained on the the Fragilistic zero-points. The calibration shifts in the retrained models are coherently propagated through training and light curve fitting.

\begin{figure*}[!h]
\includegraphics[width=19cm]{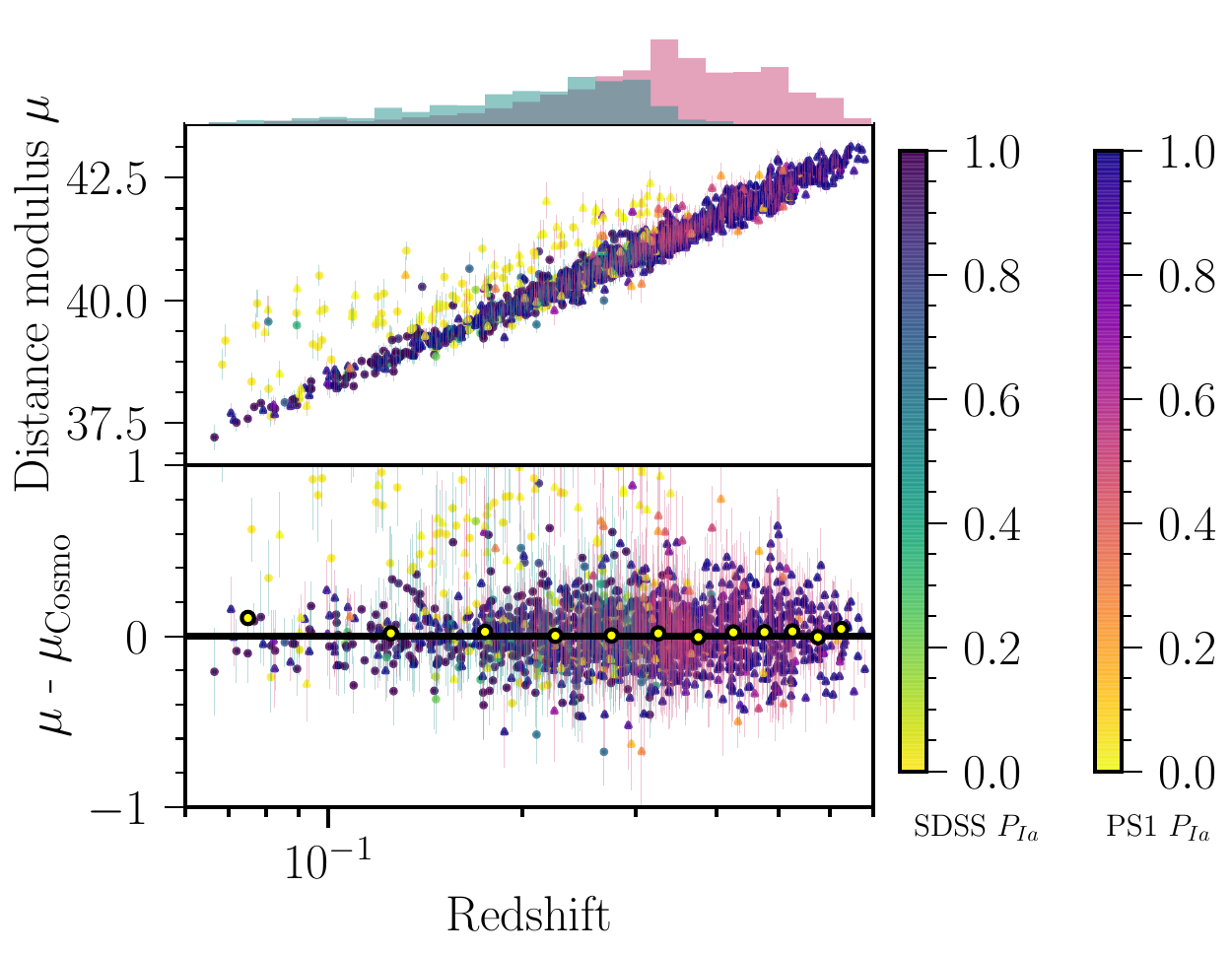}
\caption{\textbf{Top panel:} Distance Modulus $\mu$ versus redshift $z$ ("Hubble Diagram") for the Amalgame sample. SDSS (Viridis) and PS1 (Plasma, triangles) are coloured according to their classification probability $P_{\rm Ia}$ from \texttt{SuperNNova}; the Low-z and Foundation surveys are dark grey as they contain no Core collapse supernovae. \textbf{Bottom panel:} the distance modulus residual relative to a best-fit cosmological model ("Hubble Residuals"). The median Hubble Residual in bins of redshift is shown in yellow circle for reference.}
\label{fig:zHDvMU}
\end{figure*}

%%%%%%%%%%%%%%%%%%%%%%%%%%%%%%%%%%%%%%%%%%%%%%%%%%%%

\subsection{Survey Modeling} \label{sec:Inputs:subsec:surveymodel}
\uline{\textit{Purpose}}: Cosmological measurements using SNe Ia require additional followup for the acquisition of host galaxy redshifts. Inaccurate modeling of this follow-up efficiency can lead to errors in the simulated bias corrections. 

\uline{\textit{Baseline}}: In the fiducial analysis, the host galaxy detection efficiency is modeled as a function of the $r$-band magnitude of the host galaxy. The resulting redshift distribution is shown in Figure \ref{fig:basic_hists} alongside the $c$ and $x_1$ distributions.

\uline{\textit{Systematics}}: As a systematic we follow \cite{Jones19} and model the efficiency as a function of redshift (Equations 2 and 3 of \citealp{Popovic19}) for both SDSS and PS1. The new redshift efficiency is shown in Figure \ref{fig:EFF_COMP}, compared to the fiducial.

%%%%%%%%%%%%%%%%%%%%%%%%%%%%%%%%%%%%%%%%%%%%

\subsection{Intrinsic Scatter Models}
\label{sec:Inputs:subsec:scattermodel} 

\uline{\textit{Purpose}}: We use SNANA to generate catalogue-level simulations for use in bias corrections and to correct for inefficiencies in measuring intrinsic distributions of SNIa fitted parameters. Within the simulation framework, we model survey and astrophysical properties that, together, account for all sources of intrinsic scatter. Colour and stretch populations account for most of the observed intrinsic variation, but there is also a small $\sim 0.1$ mag scatter from the remaining scatter in SNIa measurements post-standardisation. This intrinsic scatter model requires a method of determining the stretch and colour populations for use in bias corrections. 

\uline{\textit{Baseline}}: To model the SNIa intrinsic scatter shown in Equation \ref{eq:errmodel} in our simulations, we use the \cite{BS20} (BS21) dust model with parameters updated in \cite{Popovic22}, hereafter P22. A more detailed analysis of the model is available in \cite{BS20} and \cite{Popovic22}. P22 simulations are used to generate BBC-4D bias corrections. As P22 does not include $x_1$ effects, the $x_1$ values are drawn from the method in \cite{Popovic21a}. 

We use the Dust2Dust program as presented in \cite{Popovic22} to derive the dust model parameters ($c_{\rm int}, R_V, E(B-V),\beta_{\rm SN}$) for the combined SDSS and PS1 sample. These model parameters (hereafter constituting the P22) are given in Appendix \ref{appendix:Dust2Dust}, and we generate our baseline bias correction simulations with these parameters. The simulations are used for the BBC-4D/BBC-BS20 bias correction method presented in \cite{Popovic21a}. A more in-depth explanation of the effects of dust on SNIa light curves is presented in BS21 and P22, but here we summarise the effects of dust on $m_B$:
\begin{equation}
    \Delta m_B = \beta_{\rm SN}c_{\rm int} + (R_V+1)E_{\rm dust} + \epsilon_{\rm noise}
    \label{eq:deltamb}
\end{equation}
and colour $c$:
\begin{equation}
    c_{\rm obs} = c_{\rm int} + E_{\rm dust} + \epsilon_{\rm noise}.
\label{eq:cobs}
\end{equation}

\uline{\textit{Systematics}}: To reflect fitting uncertainties in Dust2Dust, we follow P22 and \cite{Brout22} and draw 3 alternate realisations of model parameters from Dust2Dust as alternative bias correction models. Parent populations for $x_1$ are calculated for the SDSS+PS1 photometric sample using the methodology in \cite{Popovic21a} and the parent populations for Foundation and Low-z are taken from the same paper. 

\subsection{Core Collapse Models}\label{sec:Inputs:subsec:ccprior}

\uline{\textit{Purpose}}: We use simulations for the BEAMS method (\ref{sec:SNInfo:subsec:cc}) to model the relative luminosities and rates of Ia and non-Ia SNe. This prior is used to model the BEAMS likelihood and subsequently the weighting for the supernova in the cosmological fit.

\uline{\textit{Baseline}}: We use simulations of SDSS and PS1 with core-collapse SED inputs from \cite{Vincenzi19} to train SNN. We include Iax, 91-bg, SNII, and SNIb/SNIc simulations alongside SNIa \citep{SNIax_1, SNII_1, SNII_2, SNII_3, SNII_4, SNIbc_1, SNIbc_2, SNIbc_3, SNIbc_4}. Figure \ref{fig:MURESCC} shows the characteristic `Hubble Shoulder' arising from non-Ia in the SN sample, and that simulations match the data for Amalgame. The SDSS simulations slightly overestimate the amount of non-Ia contamination compared to the data; we choose to use this more conservative estimation over changing non-Ia rates to ensure a better match.

\uline{\textit{Systematics}}: In place of non-Ia simulations, we use the $z$-dependent polynomial expansion approximation from \cite{Hlozek12} to model non-Ia populations. This $z$-dependent polynomial adds additional parameters to the BBC fit, increasing the complexity.

%%%%%%%%%%%%%%%%%%%%%%%%%%%%%%%%%%%%%%%%%%%%%%%%%%%%

\subsection{Redshift Evolution of Nuisance Parameters }\label{sec:Inputs:subsec:nuisanceparams} 

\uline{\textit{Purpose}}: Equation \ref{eq:tripp} includes a colour-luminosity coefficient $\beta$ and a stretch-luminosity component $\alpha$ that are fit for the entire supernova sample. These nuisance parameters describe the slope of the SNIa-parameter (e.g. $c$ or $x_1$) vs. luminosity graph.

\uline{\textit{Baseline}}: We follow \cite{Brout22} and other cosmology analyses and assume that $\alpha$ and $\beta$ are constant in redshift while they are fit in the BBC process. 

\uline{\textit{Systematics}}: We include two similar systematics, allowing $\alpha$ and $\beta$ to evolve with redshift. They are modeled as 

\begin{equation}\label{eq:alphaz}
    \alpha(z) = \alpha_0 + \alpha_1 \times z
\end{equation}
and 
\begin{equation}\label{eq:betaz}
    \beta(z) = \beta_0 + \beta_1 \times z .
\end{equation}

We allow one nuisance parameter to evolve with redshift at a time.

%%%%%%%%%%%%%%%%%%%%%%%%%%%%%%%%%%%%%%%%%%%%%%%%%%%%

\subsection{Fixed Nuisance Parameters}\label{sec:Inputs:subsec:nuisancefixed}

\uline{\textit{Purpose}}: We require consistency between treatment of simulations and data in the cosmological pipeline. Unlike past analyses, which directly use nuisance parameters $\alpha$ and $\beta$ as inputs to simulations, we utilise the P22 scatter model. The process of determining the P22 scatter model, which requires an initial guess for $\alpha$ and $\beta$ in the fitting process, may result in the simulated $\alpha$ and $\beta$ values not agreeing with the data. It is unclear how this discrepancy in simulated and real $\alpha$ and $\beta$ values may affect cosmology.

\uline{\textit{Baseline}}: We let the BBC process fit for $\alpha$ and $\beta$ in the minimisation procedure, as done in historical analyses. This may result in the $\alpha$ and $\beta$ returned by the BBC process being biased.

\uline{\textit{Systematics}}: We fix the $\alpha$ and $\beta$ values to those found in a fit of the data using only redshift-based corrections to distance.

%%%%%%%%%%%%%%%%%%%%%%%%%%%%%%%%%%%%%%%%%%%%%%%%%%%%%%%%%%

%%%%%%%%%%%%%%%%%%%%%%%%%%%%%%%%%%%%%%%%%%%%%%%%%%%%%%%%%
\subsection{Validation}\label{sec:Inputs:subsec:validation}

\uline{\textit{Purpose}}: Proper analysis with simulations requires validation to ensure that we are able to recover our cosmological input parameters and not produce biases. These validation tests track biases due to BBC, light-curve fitting, and simulation inputs, but do not track assumptions made about photometry or calibration.

\uline{\textit{Baseline}}: The analysis is repeated, end to end, on 10 data-sized simulations using the P22 scatter model. 

\subsection{Other}\label{sec:Inputs:subsec:Other}

We include the following systematics as they were in \cite{Brout22}: Redshift, Peculiar Velocities, and Milky Way Reddening. These are summarised in Table \ref{tab:syst_overview}.

%%%%%%%%%%%%%%%%%%%%%%%%%%%%%%%%%%%%%%%%%%%%%%%%%%%%

\section{Results}\label{sec:Results}

Here we present the results of the Amalgame analyses.

\subsection{Nominal Cosmological Results}
\label{subsec:BaselineCosmo}

\subsubsection{Validation} 
We perform our nominal analysis on 10 simulated data-sized samples with the P22 model, with realistic distributions of SNIa parameters and core collapse contamination. Table \ref{tab:4DResults} shows a comparison of the nuisance parameters from simulated SNIa and real data. For the Amalgame sample, $\beta$ is determined from forward-modeling within the Dust2Dust program, where $\beta$ is one of the output metrics. This stands in contrast to previous analyses using G10 and C11, where $\beta$ was an independent input to the simulations. This $\beta$ discrepancy is discussed further in Section \ref{sec:Inputs:subsec:nuisancefixed}. Our $\gamma$ values for simulations and data are different, though in line with results from \cite{Brout22}. The $\sigma_{\rm grey}$ values are different as well, indicating that there still exists a relatively small amount of unexplained scatter in the data that is not being modeled by the simulations. The $\sigma_{\rm grey} = 0$ for the simulations is expected, as the BBC process is accounting for the expected scatter in the simulations. Nonetheless, we find we recover our input cosmology, with $w_{\rm reco} = -0.999 \pm 0.005$.

\subsubsection{Hubble Diagram}
The Amalgame Hubble Diagram of \NSNe\ SNIa light curves is shown in the large panel of Figure \ref{fig:zHDvMU}. It spans a redshift range of \zmin\ to \zmax. The datapoints are coloured according to their probability of being an SNIa (\ref{sec:SNInfo:subsec:cc}). The top panel is the redshift distribution of each of the constituent subsurveys. The bottom panel of Figure \ref{fig:zHDvMU} shows the residuals to the best-fit cosmology, the results of which are presented in the following subsections. 

\subsection{Consistency of Subsamples}
\label{subsec:SubsampleCosmo}

As this is the first attempt to combine two photometric samples, here we investigate the consistency across the subsamples and the full sample. The difference between the mean Hubble Residuals of the two samples is 0.0013 magnitudes, less than the errors on the Hubble Residuals themselves.

\begin{figure}[h]
\includegraphics[width=8cm]{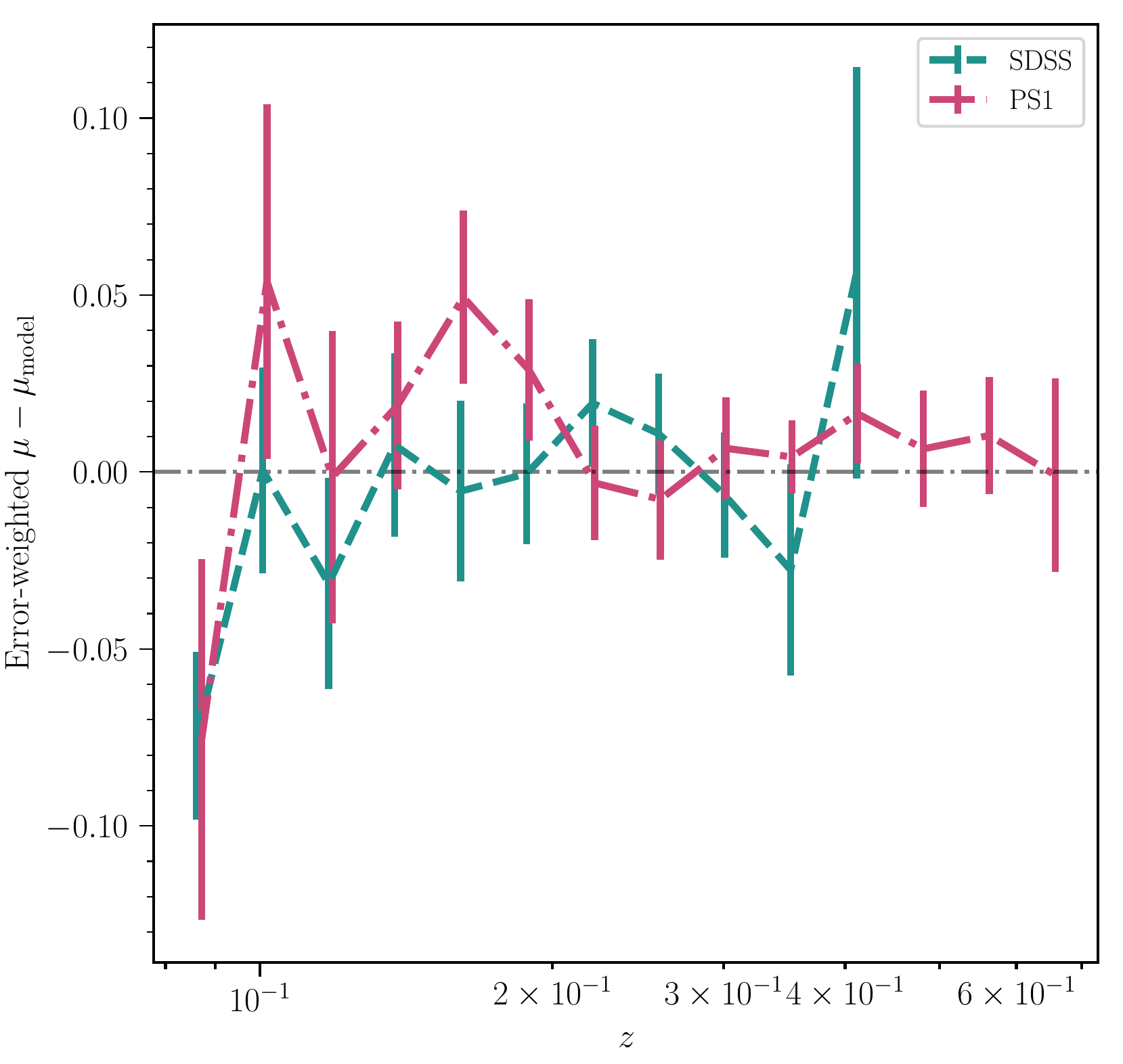}
\caption{Hubble Residuals $\mu - \mu_{\rm model}$ for subsample surveys. PS1 residuals are presented in fuschia plot and SDSS residuals in turquoise. PS1 dominates the Hubble Residuals over SDSS.}
\label{fig:MU_MUMODEL}
\end{figure}

Figure \ref{fig:MU_MUMODEL} shows the Hubble Residuals for the full Amalgame sample, split into the individual subsurveys. The SDSS and PS1 results are consistent with each other, differing by less than $1\sigma$ for any given bin. There does not present any coherent offset between the two subsamples ($<1\sigma$ confidence). 

\begin{figure}[h]
\includegraphics[width=8cm]{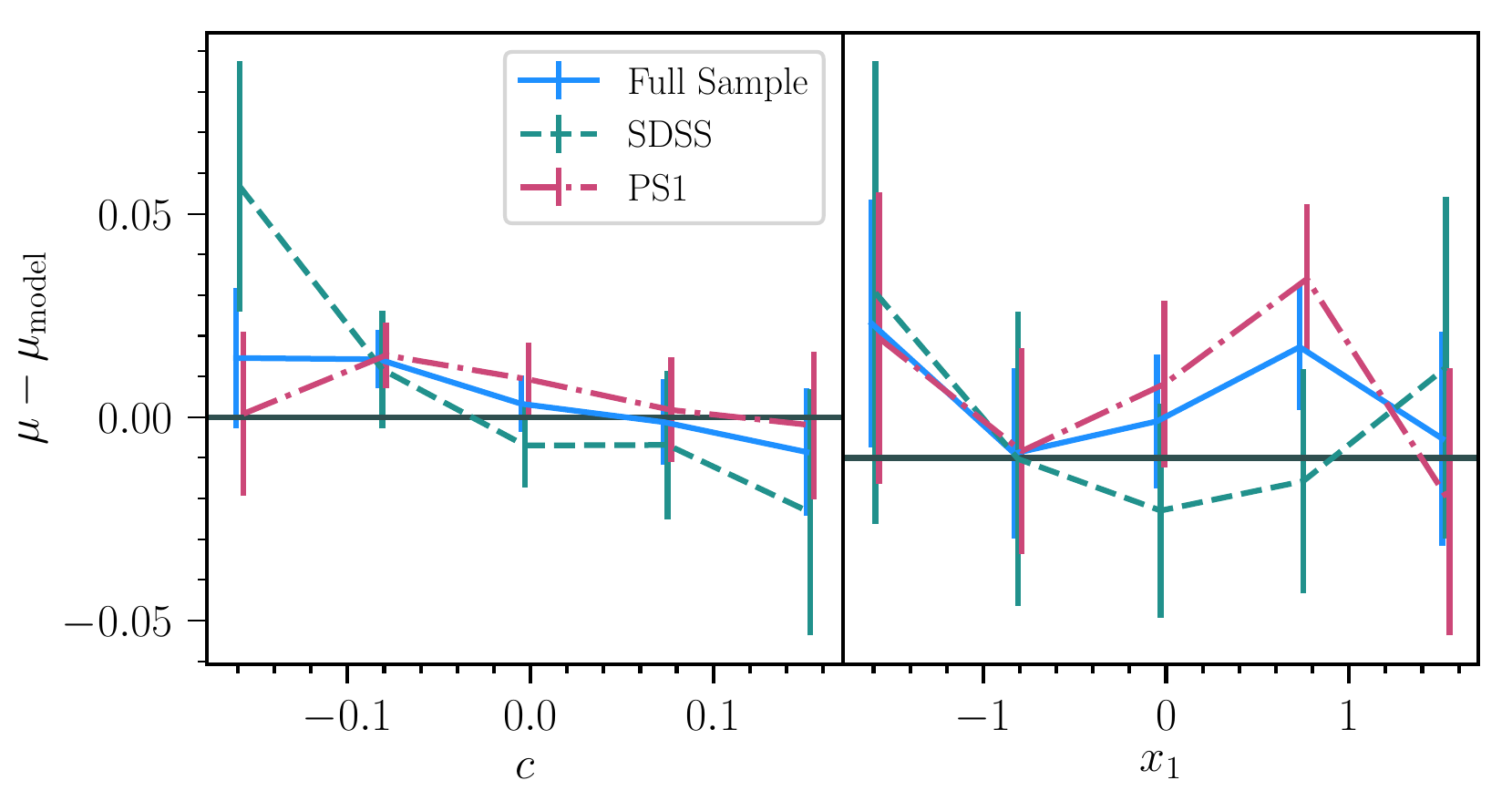}
\caption{Hubble Residuals $\mu - \mu_{\rm model}$ vs. $c$ and $x_1$ for the entire sample and subsamples. The entire sample is shown in blue; SDSS is shown in turqoise dashed line and PS1 in fuschia dotted line. The subsamples are slightly offset for visual clarity.}
\label{fig:MURESV}
\end{figure}

To further evaluate the consistency between the two photometric samples, we present the correlation of bias-corrected Hubble Residuals with $c$ and $x_1$ for the overall combined sample and the constituent subsamples in Figure \ref{fig:MURESV}. The slope of the Full Sample correlations between the Hubble Residuals and $c$ and $x_1$ (shown in blue) is consistent with zero at $1.3\sigma$ level.

\begin{figure}[h]
\includegraphics[width=8cm]{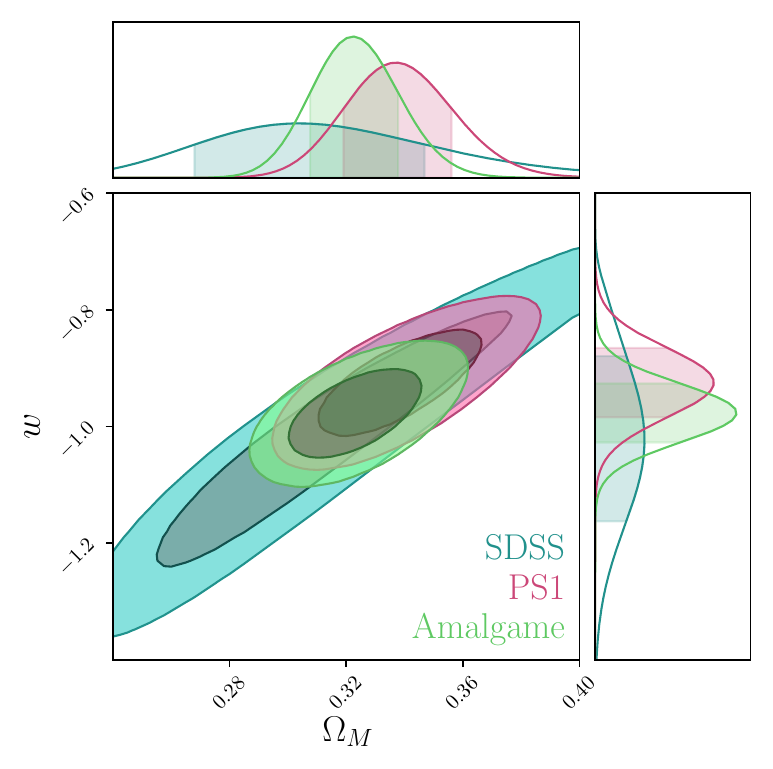}
\caption{The $w$/$\Omega_{\rm matter}$ systematic contour for the Amalgame sample calculated with \texttt{wfit} when combined with prior from \cite{Planck18}. The full sample is presented in bright green, next to PS1 in fuschia and SDSS in cyan.}
\label{fig:SubContour}
\end{figure}

This consistency in the Hubble Residuals carries through to consistency in cosmological contours. Figure \ref{fig:SubContour} shows the full Amalgame sample alongside SDSS and PS1 when combined with constraints from \cite{Planck18}, using the \texttt{wfit} minimiser. For SDSS alone, we find an $\Omega_M = 0.313 \pm 0.037$ for a Flat-$\Lambda$CDM and an $\Omega_M = 0.344 \pm 0.026$ for PS1. The full combined Amalgame constraints are discussed later, but shown here for comparison's sake between the constituent samples.

\begin{table}[]
    \centering
    \begin{tabular}{c|c|c}
    {\textbf{Amalgame}}& {\textbf{Data}} & {\textbf{Sim}} \\
    \hline
       $N_{\rm SNe}$         & \NSNEPTWENTYONEDATA & \NSNEPTWENTYONESIMDATA \\
       \hline
       $\alpha$              & \ALPHAPTWENTYONEDATA & \ALPHAPTWENTYONESIMDATA \\
       \hline
       $\beta$               & \BETAPTWENTYONEDATA & \BETAPTWENTYONESIMDATA \\
       \hline
       $\gamma_{M_{\star}}$  & \GAMMAPTWENTYONEDATA & \GAMMAPTWENTYONESIMDATA \\
       \hline
       $\sigma_{gray}$       & \SIGMAPTWENTYONEDATA & \SIGMAPTWENTYONESIMDATA \\
       \hline
       Hubble Diagram RMS                   & \RMSPTWENTYONEDATA  & \RMSPTWENTYONESIMDATA
    \end{tabular}
    \caption{{Comparison of BBC output between data and simulations}}
    \label{tab:4DResults}
\end{table}

\subsection{Constraints on Cosmological Parameters}

\begin{figure}[h]
\includegraphics[width=8cm]{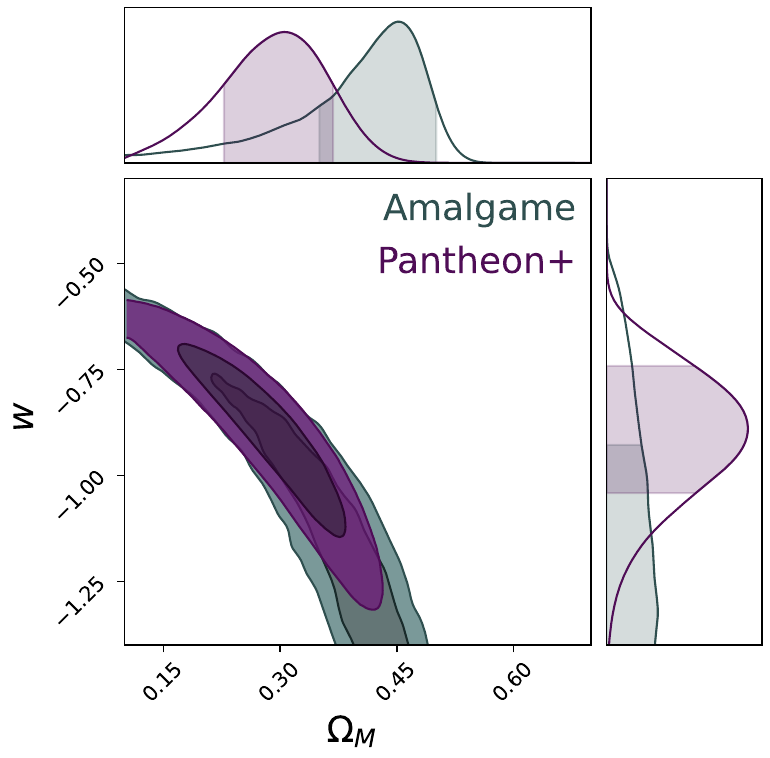}
\caption{The $w$/$\Omega_{\rm matter}$ contour for the Amalgame sample using CosmoSIS for SN-only measurement. The Amalgame is plotted in purple, alongside the SN-only results from Pantheon+ in grey.}
\label{fig:SNOnly}
\end{figure}

Here we present the cosmological constraints for the Amalgame sample using CosmoSIS. Figure \ref{fig:SNOnly} shows the SN-only cosmological contours for the Amalgame and Pantheon+ for comparison. The SN-only contours are consistent between Amalgame and Pantheon+. For the Amalgame sample combined with a CMB prior, we find $w=$ \w\ for a flat $w$CDM universe. For a flat $\Lambda$CDM universe, we find $\Omega_M = $ \omegamsn. These results are comparable with Pantheon+ at $0.334 \pm 0.018$, though with increased uncertainty from the smaller redshift range.  

\begin{table}
    \centering
    \caption{Cosmological Results}
    \label{tab:cosmology}
    \begin{tabular}{ccc}
        \hline
		Model & $\Omega_M$ & $w$  \\
            \hline
             & &  \\
            SN-Only & & \\
            \hline
            Amalgame & \omegamsn & $-1$  \\
            SDSS & $0.313 \pm 0.037$ & $-1$ \\
            PS1 & $0.344 \pm 0.026$ & $-1$ \\
             & &  \\
		\textcolor{gray}{Pantheon+} & \textcolor{gray}{$0.334 \pm 0.018$} & $-1$ \\
             & &  \\
            with CMB & &  \\
		\hline
            Amalgame & \omegam & \w\  \\
             & &  \\
            \textcolor{gray}{Pantheon+} & \textcolor{gray}{$0.325^{+0.010}_{-0.008}$} & \textcolor{gray}{$-0.982^{+0.022}_{-0.038}$} \\

    \end{tabular}
\end{table}

\begin{figure}[h]
\includegraphics[width=8cm]{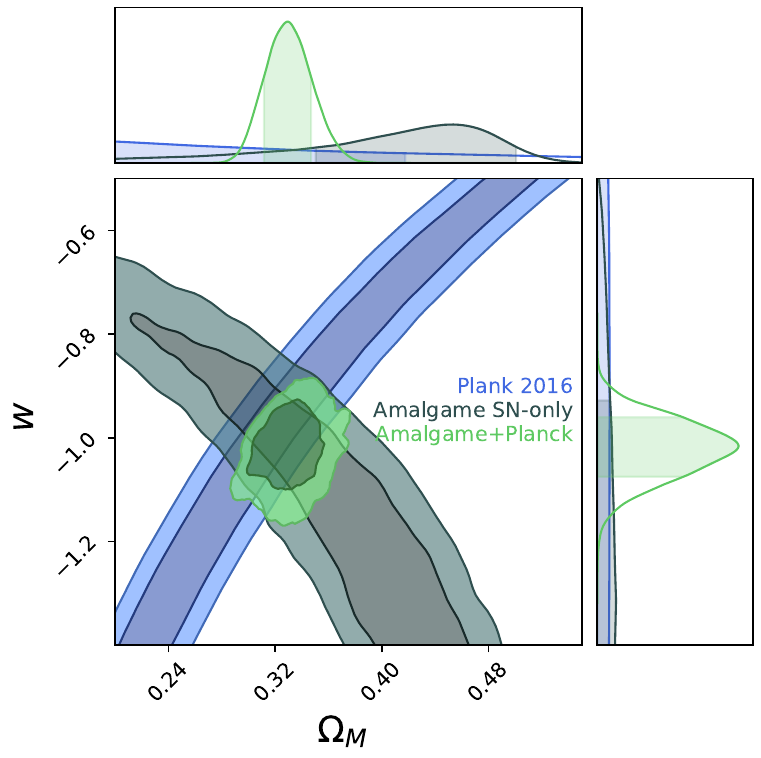}
\caption{The $w$/$\Omega_{\rm matter}$ contour for the Amalgame sample with a prior from the \cite{Planck18} measurement using Cosmosis. The SN-only results from Amalgame are shown in purple, Planck 2020 in blue, and the combined measurement in bright green.}
\label{fig:FullContour}
\end{figure}

Figure \ref{fig:FullContour} shows the $w$/$\Omega_{\rm matter}$ constraints when combined with the results from \cite{Planck18} from the full Amalgame sample. 

Table \ref{tab:cosmology} displays a summary of the cosmological measurements from Amalgame, $\Omega_m$ and $w$, as SN-only measurements and with CMB results from \cite{Planck18}.

\subsection{Systematic Uncertainties}

\begin{figure*}[h]
\includegraphics[width=16cm]{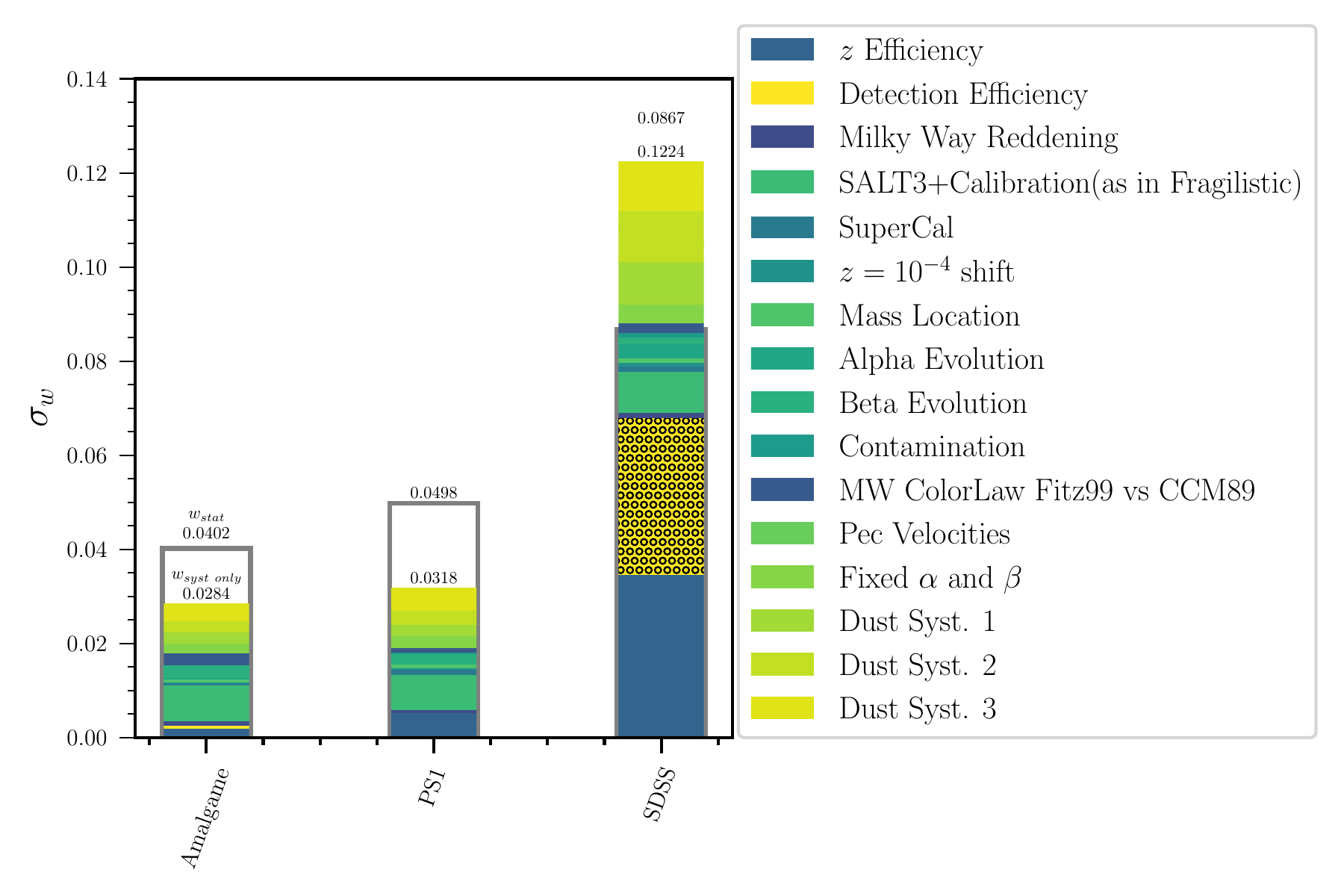}
\caption{Error Budget comparison for the subsample combinations presented in this paper. The full sample is presented first, followed by PS1 and SDSS. PS1 has a similar uncertainty to Amalgame, but SDSS is considerably larger; the systematic uncertainty for SDSS is larger than the statistical uncertainty; the gray statistical uncertainty can be seen behind the systematic uncertainty. For Amalgame, we find $w = $\w.}
\label{fig:ERR_BUDGETS}
\end{figure*}

Table \ref{tab:syst_overview} provides a comprehensive overview of the sources of systematic uncertainty for the Amalgame analysis. These uncertainties, calculated with \texttt{wfit}, and their relative contribution to the overall $w$ systematic uncertainty, are displayed in Figure \ref{fig:ERR_BUDGETS}. When comparing the overall Amalgame $w$-uncertainties with those of SDSS and PS1 (seen in Figure \ref{fig:ERR_BUDGETS}), the statistical $w$-uncertainty increases as expected, but alongside the statistical, the systematic $w$-uncertainty also increases. This indicates that some systematics depend on the sample size used to set a constraint, and that future, larger surveys using covariance matrix measurements of cosmology may enjoy reductions in systematic uncertainty. 

\begin{table*}
    \centering
        \caption{Sources of Uncertainty}
    \begin{tabular}{p{3.9cm}p{3.9cm}p{3.9cm}p{0.9cm}p{0.9cm}p{0.9cm}p{1.1cm}} %{lllcccc}
Sec. and description & Baseline & Systematic & Scale & $\sigma_{w,\mathrm{syst}}$ & $\frac{\sigma_{w,\mathrm{syst}}}{\sigma_{w,\mathrm{stat}}}$ & $\Delta w_{\mathrm{syst}}$ $^{\dag}$\\

\hline
\multicolumn{4}{l}{\textbf{Total Statistical}} & N/A & 1.00 & \DELTAWALL \\
\multicolumn{4}{l}{\textbf{Total Systematic} (Unbinned Covariance Matrix Approach)} & \SIGSYS & \FRACSYSALL & N/A \\
%\multicolumn{4}{l}{Total Systematic (Binned Covariance Matrix Approach )} & (0.028) & (0.84) & 0.014 \\

\hline
\multicolumn{7}{l}{\textbf{Data}} \\ %                                                        scale sig frac delta
- Redshift shift (\ref{sec:Inputs:subsec:Other}) & nominal redshift & $\Delta z = 10^{-4}$ & 1 & \SIGSYSZSHIFT & \FRACSYSZSHIFT & \DELTAWZSHIFT \\
- Peculiar velocities (\ref{sec:Inputs:subsec:Other}) & 2M$++$ velocity map & 2MRS velocity map \footnote{Both from \cite{Peterson21}} & 1 & \SIGSYSVPEC  & \FRACSYSVPEC & \DELTAWVPEC \\
- Host Galaxy Properties (\ref{sec:Inputs:subsec:hostgal}) & Mass step at $M_{\star} = 10$ & Mass step at $M_{\star} = 10.2$ & 1 & \SIGSYSMASSLOC &  \FRACSYSMASSLOC & \DELTAWMASSLOC \\
%- Classification algorithm (\ref{sec:Inputs:subsec:ccprior}) & SuperNNova  &  SCONE & 1 & ongoing & ongoing & ongoing \\
- $\alpha$ Evolution (\ref{sec:Inputs:subsec:nuisanceparams}) & constant $\alpha$ & $\alpha(z) = \alpha_{0} + \alpha_{1}\times z$ & 1 & \SIGSYSALPHAEVOL & \FRACSYSALPHAEVOL & \DELTAWALPHAEVOL \\
- $\beta$ Evolution (\ref{sec:Inputs:subsec:nuisanceparams}) & constant $\beta$ & $\beta(z) = \beta_{0} + \beta_{1}\times z$ & 1 & \SIGSYSBETAEVOL & \FRACSYSBETAEVOL & \DELTAWBETAEVOL \\
- Not floating $\alpha$ and $\beta$ in BBC Fit (\ref{sec:Inputs:subsec:nuisancefixed}) & Floated $\alpha$ and $\beta$ & Fixed $\alpha$ and $\beta$ values & 1 & \SIGSYSFIXED & \FRACSYSFIXED & \DELTAWFIXED \\
\hline

\multicolumn{7}{l}{\textbf{Calibration and Light-curve modelling}} \\
- HST Calspec (\ref{sec:Inputs:subsec:calibration}) & Calspec 2020 Update & 5 mmag/7000\AA & 3 & \SIGSYSCALSPEC & \FRACSYSCALSPEC & \DELTAWCALSPEC \\
- SALT3 surfaces (\ref{sec:Inputs:subsec:salt3}) $\&$ ZP & SALT3 trained on fragilistic best-fit of K21 & 10 covariance realizations  & 1/3 & \SIGSYSCALSALT & \FRACSYSCALSALT & \DELTAWCALSALT \\
- MW scaling (\ref{sec:Inputs:subsec:Other}) & \cite{Schlafly11} & 5$\%$ scaling & 1 &  \SIGSYSMWEBV & \FRACSYSMWEBV & \DELTAWMWEBV \\
- MW colour law (\ref{sec:Inputs:subsec:Other}) & $R_V$=3.1 and F99 & $R_V$=3.0 and CCM & 1/3 & \SIGSYSCOLORLAW & \FRACSYSCOLORLAW & \DELTAWCOLORLAW \\
\hline

\multicolumn{7}{l}{\textbf{Simulations}}\\
- Survey Modelling $\epsilon_{z}^{\mathrm{spec}}$ [X] (\ref{sec:Inputs:subsec:surveymodel}) & r-band magnitude efficiency & $z$-based efficiency & 1 & \SIGSYSZEFF & \FRACSYSZEFF & \DELTAWZEFF \\
- Intrinsic scatter model (\ref{sec:Inputs:subsec:scattermodel}) &  Dust model parameters from Appendix Table \ref{tab:dust2dust} & Separate dust parameters within $1\sigma$ &  &  &  &  \\

 & & Sys 1 & 1/$\sqrt{3}$ & \SIGSYSBCONE & \FRACSYSBCONE & \DELTAWBCONE \\
 & & Sys 2 & 1/$\sqrt{3}$ & \SIGSYSBCTWO & \FRACSYSBCTWO & \DELTAWBCTWO \\
 & & Sys 3 & 1/$\sqrt{3}$ & \SIGSYSBCTHREE & \FRACSYSBCTHREE & \DELTAWBCTHREE \\
- Core-collapse SN prior (\ref{sec:Inputs:subsec:ccprior}) & \cite{Vincenzi19} CC Templates & Polynomial fit as in Holzek 2012 & 1 & \SIGSYSSNN & \FRACSYSSNN & \DELTAWSNN \\
\hline
    \end{tabular}
    ${\dag}$ {Shift in $w$ when including when including ONLY this systematic}
\label{tab:syst_overview}
\end{table*}

\begin{figure}[h]
\includegraphics[width=8cm]{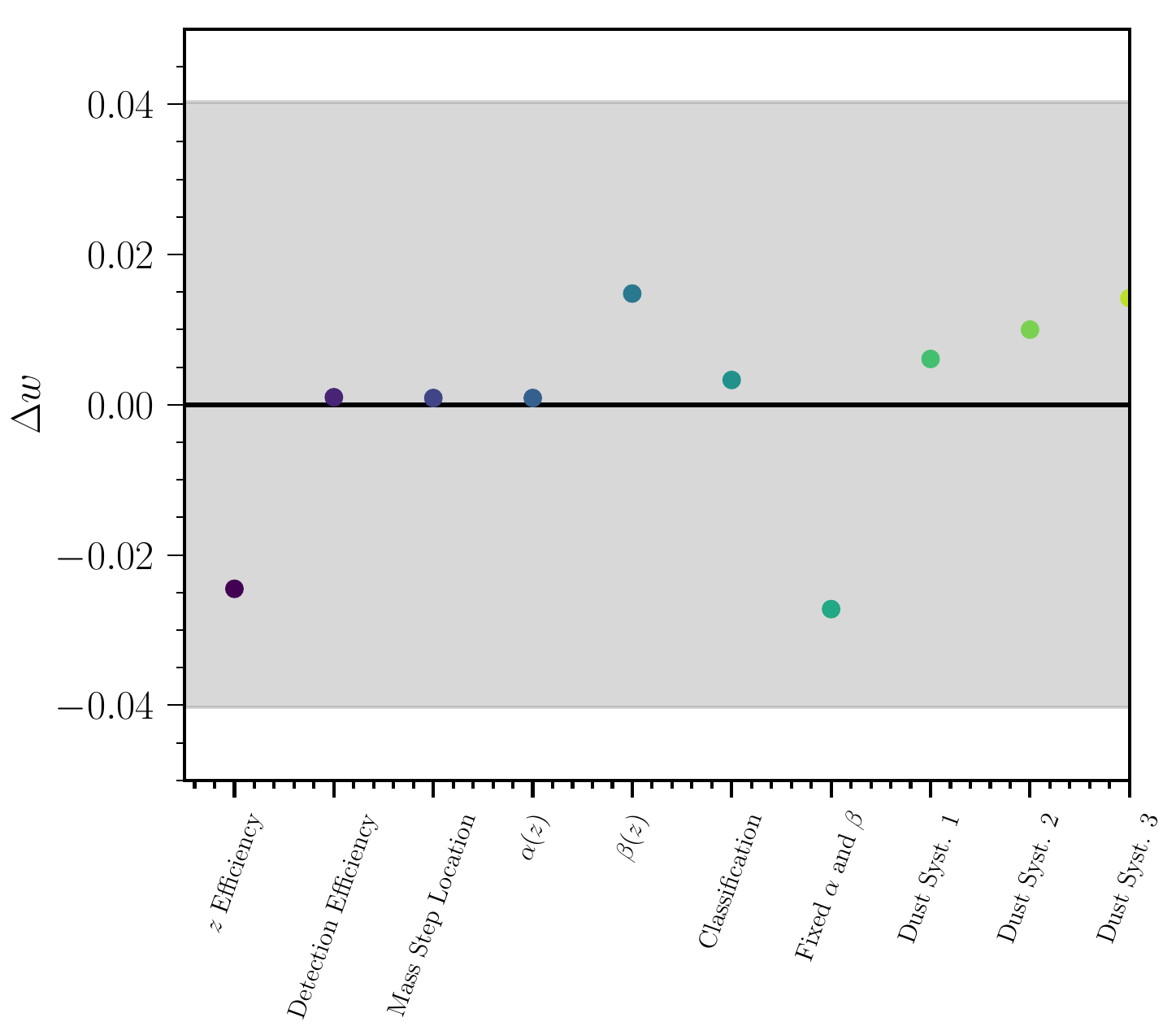}
\caption{Changes in measured $w$ between systematics for the fiducial Amalgame sample calculated with \texttt{wfit}. Grey fill is the statistical uncertainty from the fiducial analysis. These measurements do not include the Covariance Matrix from Section \ref{sec:SNInfo:subsec:covariance}.}
\label{fig:DELTAW}
\end{figure}

\section{Discussion}\label{sec:Discussion}

Overall, the individual systematics are subdominant to the statistical uncertainty for the Amalgame sample, and for the most part on the order of $<1\%$ in $w$. The largest systematic contribution remains the calibration and SALT3 surfaces as detailed in \cite{fragilistic} and Section \ref{sec:Inputs:subsec:salt3}. This systematic uncertainty contribution, about $\sim 40 \%$ of the statistical uncertainty, is comparable to that of Pantheon+, which found a calibration and SALT3 uncertainty of $38 \%$ of the statistical uncertainty.

Those systematics shared with Pantheon+ are largely similar, instead, we shall mostly focus on those systematics unique to this analysis or that we include significant updates to (dust-based scatter modeling): 

\begin{itemize}
    \item Redshift Evolution of $\alpha$ and $\beta$
    \item Fixing $\alpha$ and $\beta$ in the BBC fit
    \item Host Galaxy Follow-up Efficiency
    \item Intrinsic Scatter Model
    \item Core Collapse modeling
    \item SDSS Detection Efficiency
\end{itemize}

\textbf{$\alpha(z)$ and $\beta(z)$} \\
We allow the stretch-luminosity coefficient $\alpha$ and the colour-luminosity coefficient $\beta$ to change with $z$. We find no evidence that $\alpha$ that evolves with redshift, nor any noteworthy systematic uncertainty associated with such an evolution. However, we do find evidence of an evolving $\beta$ with redshift at a $4\sigma$. While the $\beta(z)$ systematic has a greater contribution to the overall systematic uncertainty - \SIGSYSBETAEVOL\ - it is still $\times 6$ smaller than the statistical uncertainty. 

\textbf{Fixed $\alpha$ and $\beta$} \\ 
We find a noticeable difference in the fitted $\alpha$ and $\beta$ values between the BBC-4D recovered values and those provided by Dust2Dust. The differences between the BBC-4D recovered values and the fixed values are $\Delta \alpha \sim 0.03$ and $\Delta \beta \sim 0.4$, outside statistical error. However, the resulting $\Delta w $ from fixing the nuisance parameters rather than fitting is $\Delta w =~\DELTAWFIXED$, and the systematic uncertainty is similarly small at $0.005$.

\textbf{Follow-up efficiency} \\ 
Redshift determination for photometric surveys is a function of the host-galaxy properties, not the supernovae. This systematic is $\sigma_{w, \rm{syst}} =  \SIGSYSZEFF$, and $\Delta w  =\DELTAWZEFF$. For the full Amalgame sample, this systematic makes no significant impact on cosmology. This is a sizeable systematic for SDSS alone, however. The fiducial case of using the $r$-band magnitude of the host galaxy, rather than a statistical redshift efficiency, is a more realistic approach. We find that this systematic is very well constrained, not meriting further research.

\textbf{Intrinsic Scatter Model} \\
The three dust systematics result in a combined systematic uncertainty of $\sim 0.018$, one of the larger systematic contributions. The individual contributions are approximately equal, and there does not appear to be an obvious correlation between the systematic uncertainty contribution and the $\Delta w$. 

\textbf{Core Collapse Modeling} \\
The choice of using a simulated core collapse SNe sample as the prior for BEAMS versus using the \cite{Hlozek12} analytical solution results in a negligible difference in $\Delta w$ and $\sigma_w$. While the simulated set is a more realistic approximation of the core collapse population, the CC contamination as identified by SNN effectively marginalises issues due to contamination. Noticeably, we find that non-Ia contamination represents a small portion of our systematic uncertainties.

\textbf{SDSS Detection Efficiency} \\
The change in the SDSS detection efficiency is negligible for the full Amalgame sample, though alongside the host galaxy follow up efficiency it is one of the largest systematic uncertainties for SDSS alone. Given the lower redshift range of SDSS $(0.06 < z < 0.35)$, it is expected that systematics that affect the observed redshift distribution have a large impact on the measured cosmology. 

\begin{figure}[h]
\includegraphics[width=8cm]{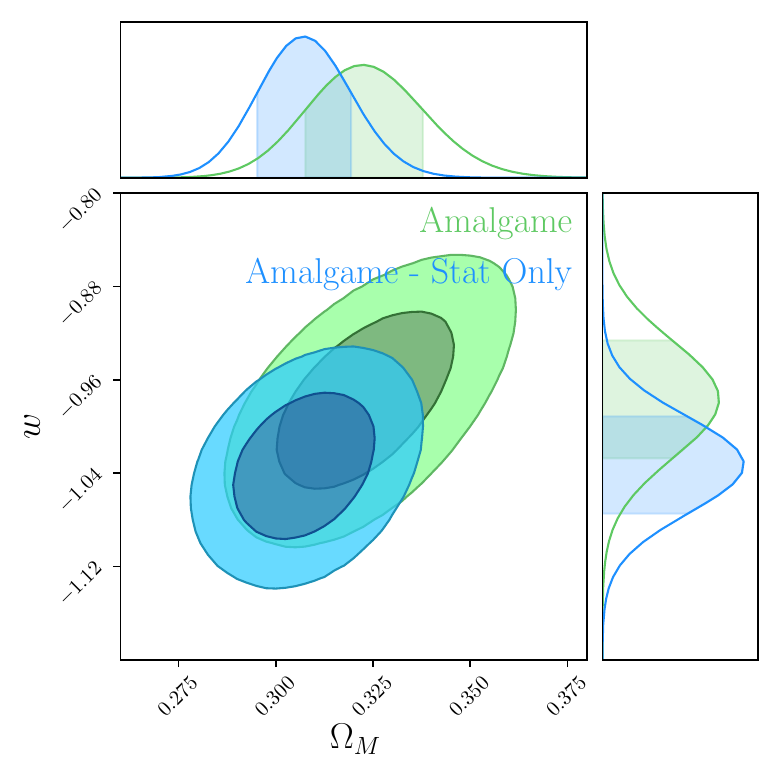}
\caption{The Statistical-Only Amalgame $\Omega_M$/$w$ contour in blue plotted next to the full Amalgame Statistics+Systematics contour in bright green. Both are calculated with \texttt{wfit}.}
\label{fig:StatSyst}
\end{figure}

Figure \ref{fig:DELTAW} shows $\Delta w$ from each systematic evaluated separately, and without using $C_{\rm syst}$ from Section \ref{sec:SNInfo:subsec:covariance}. The statistical-only $\Delta w \approx 0.04$, shown in grey fill. These represent the largest potential $\Delta w$ for this analysis, and are all within the reported statistical uncertainty.

The statistical-only constraints are shown alongside the systematic+statistical constraints for Amalgame in Figure \ref{fig:StatSyst} when combined with a CMB prior from \cite{Planck18}. These contours are generated using the \texttt{wfit} program. Compared to stat-only results, we see a shift towards higher $\Omega_M$ and lower $w$ when including systematics.

\section{Conclusion}\label{sec:Conclusion}

In this paper, we have presented the first combination of separate photometric SN samples, and provide a cosmological measurement from the combined sample. These two photometric samples are consistent at both the Hubble Residual level, and within their fitted cosmologies. Alongside these results, we have introduced updates to the bias corrections methodology introduced in \cite{Popovic21a} and \cite{Brout22}, as well as the dust model methodology from BS21 and \cite{Popovic22}. We release the data and analysis tools to facilitate public work and discussion.

When using only SN for our measurement, we find $\Omega_m = $ \omegamsn, consistent with the results of Pantheon+. Our result for the dark energy equation-of-state, combined with the CMB, is $w = $ \w, consistent with the cosmological constant of $w = -1$ and constraints from other SN samples. This consistency is in spite of the three major differences between recent cosmological measurements using SNe Ia  \citep{Scolnic18, Brout22, DES3YR} and Amalgame: the use of a photometric sample, rather than spectroscopic, for the analysis; the use of dust modeling to explain SNIa intrinsic scatter; and the lack of a low redshift anchor. Our analysis is most similar to that from \cite{Jones18}, though that focuses on one photometric sample and includes a spectroscopically confirmed low-redshift anchor.  That analysis finds a $w = -0.989 \pm 0.057$.

This paper is a stepping stone for the next generation of cosmological measurements with SNIa between now and the dawn of the Legacy Survey of Space and Time  and the Roman Space Telescope; using combined photometric samples including PS1, SDSS, and DES with an improved low-redshift sample from the Zwicky Transient Facility and the Young Supernova Experiment.

\section{Acknowledgments}

DS and DB is supported by Department of Energy grant DESC0010007, the David and Lucile Packard Foundation, the Templeton Foundation and Sloan Foundation. 
M.V was partly supported by NASA through the NASA Hubble Fellowship grant HST-HF2-51546.001-A awarded by the Space Telescope Science Institute, which is operated by the Association of Universities for Research in Astronomy, Incorporated, under NASA contract NAS5-26555.
We acknowledge the University of Chicago’s Research Computing Center for their support of this work.

\newpage

\begin{appendix}

\section{A. Rederiving Host Galaxy Masses}
\label{sec:HostMass}
We use photometric data for the host galaxies, measured in $grizY$ for PS1 and $ugriz$ for SDSS. The host galaxy data are corrected for Milky Way extinction correction using colour excess $E(B-V)$ values from Schlegel et al. 1998 and a Fitzpatrick reddening law (Fitzpatrick 1999).

We then estimate stellar masses for the SN host galaxies. In brief, we use the PEGASE.2 spectral synthesis code (Fioc \& Rocca-Volmerange 1997; Le Borgne \& Rocca-Volmerange 2002) to calculate the SED of a galaxy as a function of time, using smooth, exponentially declining star formation histories (SFHs). Each SED is calculated at 102 time-steps from 0 to 14 Gyr, with the standard PEGASE.2 prescription for nebular emission included. We use a Kroupa (2001) initial mass function (IMF), and seven foreground dust screens with a colour excess ranging from 0.0 to 0.30 mag in steps of 0.05 mag to mimic the effect of foreground dust. For each host galaxy, the fluxes of each model SED at the redshift of the SN in the relevant filters are calculated and the best-fitting template located using a least-squares approach. To ensure consistency with our assumed cosmological model, we enforce that the age of the best-fitting template must be less than the age of the Universe at the redshift of the SN. The stellar mass is then estimated from the best-fitting SED.

We use a Monte Carlo approach to estimate the statistical uncertainties in our stellar masses. For each galaxy, we perform 1000 random realizations of the observed galaxy data, drawing new \lq observations\rq\ randomly from a normal distribution, and repeating the procedure described above. The quoted uncertainties on the best-fitting parameters are the standard deviation of the best-fitting parameters over all realizations.

\section{B. Updates to Dust2Dust}
\label{appendix:Dust2Dust} 

Proper modeling of dust parameters for photometric samples requires improvements to the methodology introduced in \cite{Popovic22}. The forward modeling of dust parameters was done with use of a program named Dust2Dust; here we detail updates to the Dust2Dust approach to incorporate the presence of core collapse supernovae in the Dust2Dust fit:
\begin{itemize}
    \item  We fix $\alpha$ and $\beta$ in the Dust2Dust fit, instead of floating them in the SALT2mu minimisation.
    \item  We institute a cut on the probability of being a Ia from our classifiers $(PIa > 0.5)$. 
    \item  We do not include core collapse supernovae in the bounded simulations for Dust2Dust.
    \item  We utilise BEAMS with the \cite{Hlozek12} core-collapse prior.
\end{itemize}

In P22, the $\beta$ ($\beta_{\rm obs}$ in P22) parameter is used as a metric criteria (Section 4.1 of P22). However, this presents issues with the colour vs. Hubble Residuals metric, which is calculated by subtracting out observed distances from the best fit cosmology for both data and simulations. When $\beta_{\rm obs}$ is floated, the resulting $\beta_{\rm obs}$ for the data and simulations are not guaranteed to be the same, therefore resulting in a bias in distance moduli that is not constrained by the more impactful colour vs. Hubble Residual metric (Table 6 of P22). 

To fix this issue, the $\beta_{\rm obs}$ of the data is measured via a SALT2mu fit (\cite{Kessler16} find negligible differences between $\beta$ determinations with 1D bias corrections and none) and then used as the nominal $\beta_{\rm obs}$ for both the simulations and the data. While this approach does remove $\beta_{\rm obs}$ as a metric criteria, it is not an effective constraint on dust model parameters, contributing little to the overall $\chi^2$.

\begin{figure}[h]
\includegraphics[width=8cm]{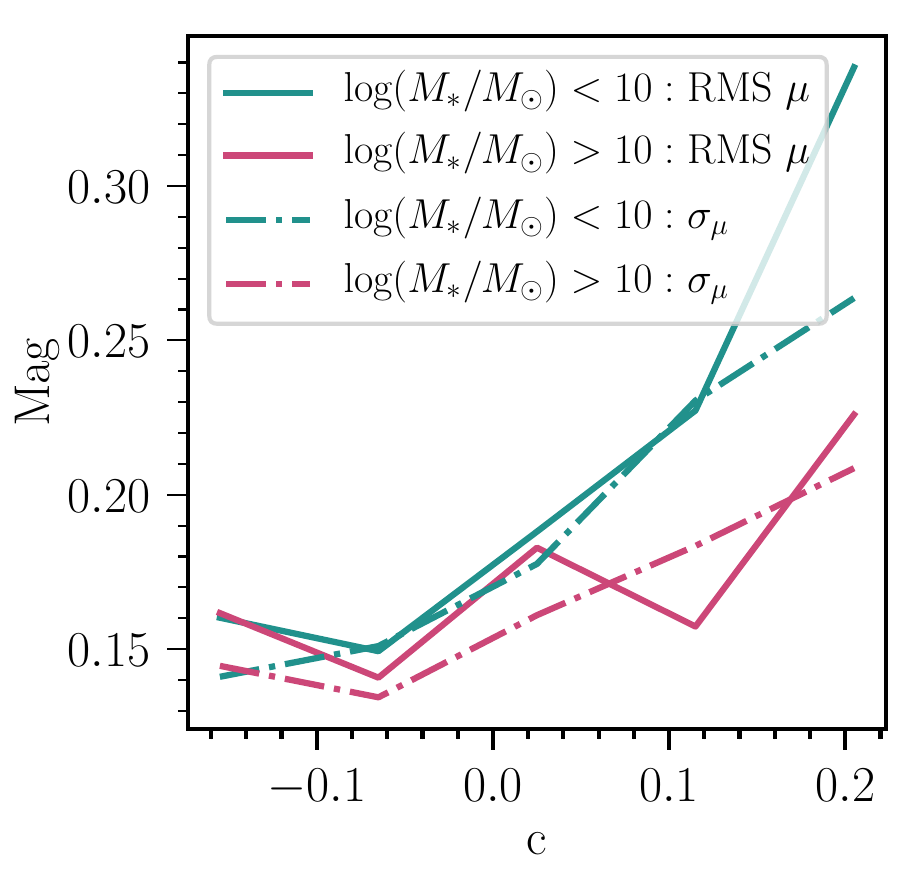}
\caption{A comparison of the Root Mean Square of the Hubble residuals (solid) with the median distance modulus error (dash dotted) for the data as a function of colour. This is shown for high and low mass host galaxies.}
\label{fig:scatter_error}
\end{figure}

To mitigate core collapse supernovae from contaminating the Dust2Dust fits, we implement a probability cut on the likelihood of being a Type Ia as assigned by our classifiers. The inclusion of obvious $(P_{Ia} < 0.5)$ core collapse supernova impacts all of the Dust2Dust criteria, and in particular the colour distribution, Hubble Residual scatter, $\beta_{\rm obs}$, and $\sigma_{\rm int}$ metrics are not able to be mitigated with the inclusion of BEAMS. Our nominal probability cut is an SNN-likelihood of greater than 0.5. As a quick test, we use our nominal dust model parameters with a probability cut of $P_{Ia} > 0.9$ to compare the results, and find a $\Delta \chi^2 = 11$.

We make the assumption that the probability cut removes enough non-Ia that the effects of not simulating core-collapse supernovae into the bounded simulations that comprise Dust2Dust are negligible. This approach has the added benefit of allowing us to use \cite{Vincenzi19} core collapse models, which include accurate dust information for core collapse supernovae.

Finally, we utilise the BEAMS methodology to mitigate the effect of any remaining core collapse supernovae on our colour vs. Hubble Residual metric. We present the best-fit Dust2Dust Amalgame samples along with a breakdown of their $\chi^2$ in Table \ref{tab:dust2dust}.

\section{C. Updates to $\sigma_{\rm int}$}
\label{appendix:sigint}

Equations \ref{eq:tripperr} and \ref{eq:errmodel} in Section \ref{sec:SNInfo:subsec:distance} detail the calculation of $\sigma_{\rm int}$ as prescribed in \cite{Brout22}. Here we detail a small correction to the calculation as done in \cite{Brout22}. 

When calculating $\mu$ for SNIa in the biasCor files, a $\beta$ must be assumed. \cite{Brout22} follow \cite{Popovic21a} in using the input $\beta_{\rm SALT}$ when calculating $\mu$; however, $\mu_{\rm ERR}$ was calculated with $\beta_{\rm SN}$ due to an overlooked variable. Calculating $\mu$ and $\mu_{\rm ERR}$ with inconsistent $\beta$ values incorrectly increased the calculated $\sigma_{\rm int}$ values. This has since been fixed to use a consistent $\beta_{\rm SALT}$.

\begin{table}[]
        \label{tab:dust2dust}
        \begin{tabular}{c|c}
        Parameter & Value \\
        \hline
        $c_{\rm int}$ & -0.074 \\
        $c_{\rm std}$ &  0.055 \\ 
        High mass $\overline{R_V}$ & 3.17 \\
        High mass $\sigma_{RV}$ & 1.23 \\ 
        Low mass $\overline{R_V}$ & 1.71 \\
        Low mass $\sigma_{RV}$ & 0.82 \\ 
        Low $z$, low mass $\tau_{EBV}$ & 0.13 \\
        Low $z$, high mass $\tau_{EBV}$ & 0.12 \\
        High $z$, low mass $\tau_{EBV}$ & 0.11 \\
        High $z$, high mass $\tau_{EBV}$ & 0.13 \\
        $\overline{\beta}_{\rm int}$ & 1.50 \\
        $\sigma_{\beta \rm{int}}$ & 0.29 
        \end{tabular}
\quad
        \begin{tabular}{c|c}
            Metric Criteria & $\chi^2$ \\
            \hline
            $\chi^2_c$ & 5.3 \\
            $\chi^{2}_{\mu_{\rm res}, \rm high}$ & 27.3 \\
            $\chi^{2}_{\mu_{\rm res}, \rm low}$ & 13.1 \\
            $\chi^{2}_{\sigma_{\rm r}, \rm high}$ & 62.3 \\
            $\chi^{2}_{\sigma_{\rm r}, \rm low}$ & 20.9 \\
            $\chi^2_{\beta_{\rm int}}$ & N/A  \\
            $\chi^2_{\sigma_{\rm int}}$ & 0.64
        \end{tabular}
\caption{P22 Model Parameters and Criteria Fit for this paper.}
\end{table}

\section{Sims and Data}
\label{sec:SimsNData}

Here we present the data-sim overlays for the SDSS and PS1 surveys. There is good agreement between the data and simulations. The redshift distribution for SDSS is slightly discrepant; this discrepancy is addressed in the efficiency modeling systematic in Section \ref{sec:Inputs:subsec:surveymodel}.

\begin{figure*}[h]
\includegraphics[width=19cm]{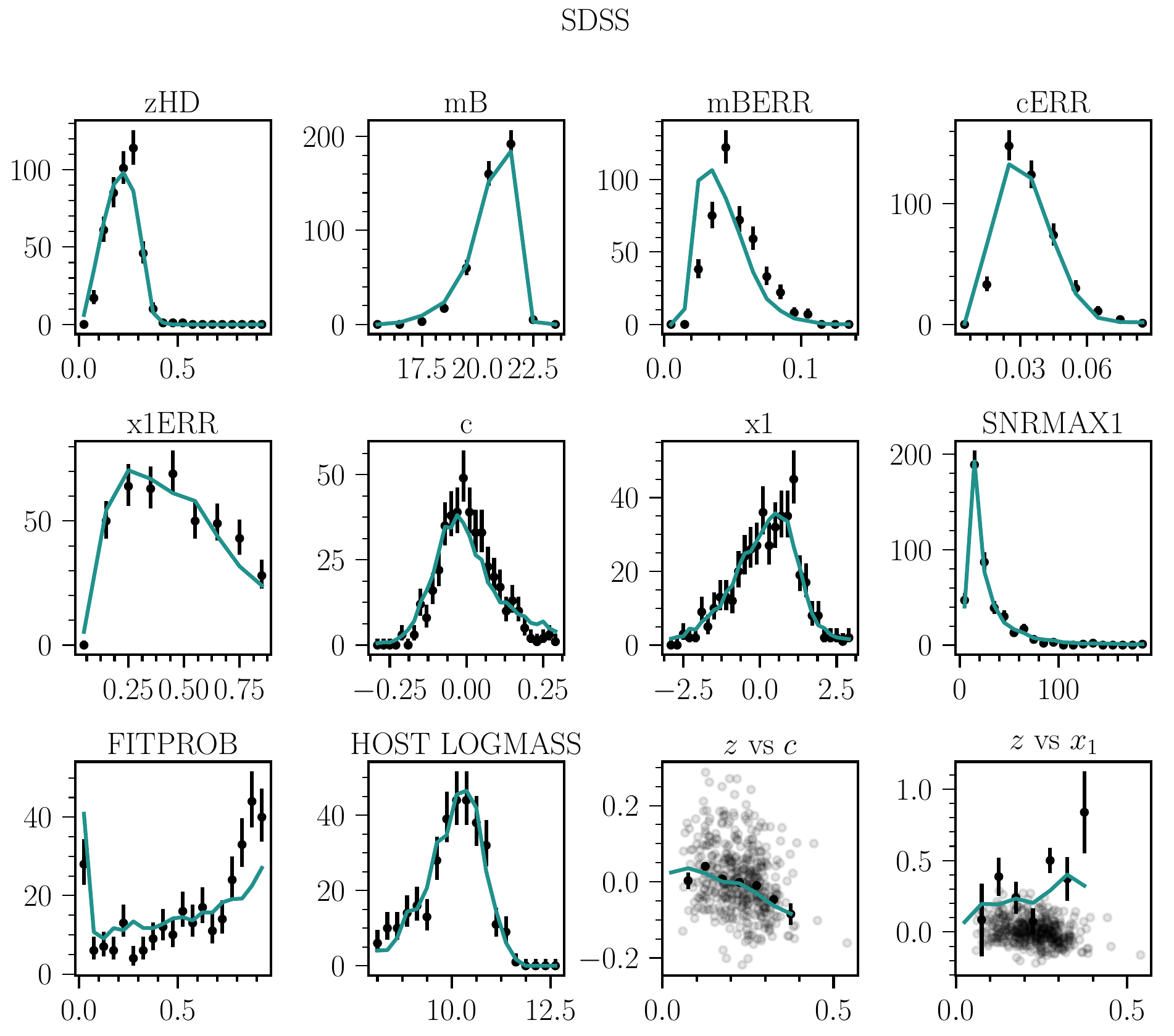}
\caption{Metrics for goodness-of-fit for SDSS. }
\label{fig:CompareSDSS}
\end{figure*}

\begin{figure*}[h]
\includegraphics[width=19cm]{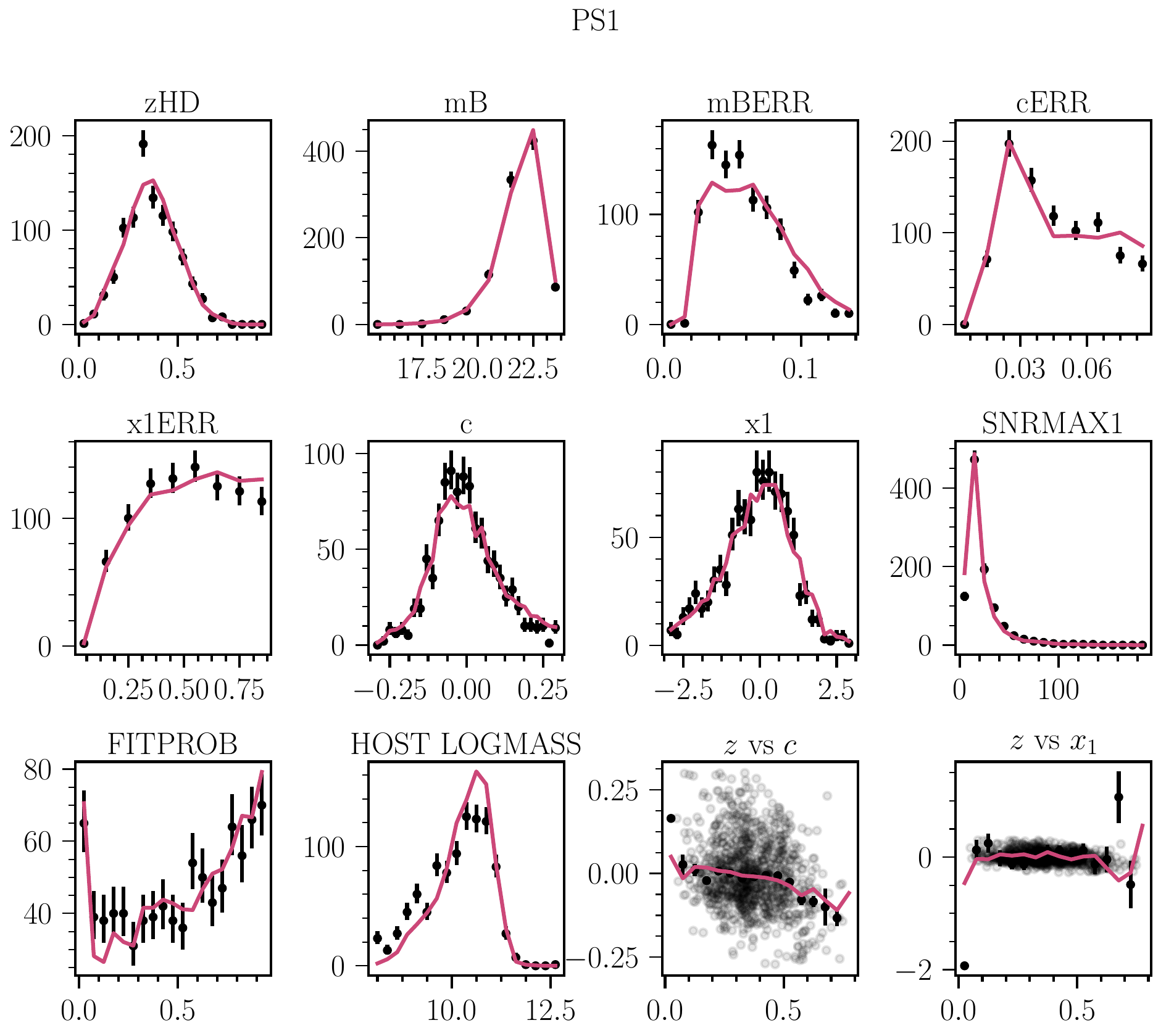}
\caption{Metrics for goodness-of-fit for PS1. }
\label{fig:ComparePS1}
\end{figure*}

\end{appendix}

\bibliographystyle{mne2.bst}

\bibliography{research2.bib}

\begin{thebibliography}{73}
\providecommand{\natexlab}[1]{#1}
\providecommand{\url}[1]{\texttt{#1}}
\providecommand{\urlprefix}{URL }
\providecommand{\eprint}[1][]{\url{#1}}

\bibitem[{{Abbott} et~al.(2019){Abbott}, {Allam} et~al.}]{DES3YR}
{Abbott}, T.~M.~C., {Allam}, S., {Andersen}, P., et~al., 2019, \apjl, 872, 2,
  L30, \eprint arXiv:{1811.02374}

\bibitem[{{Alam} et~al.(2017){Alam}, {Ata} et~al.}]{Alam17}
{Alam}, S., {Ata}, M., {Bailey}, S., et~al., 2017, \mnras, 470, 2617, \eprint
  arXiv:{1607.03155}

\bibitem[{{Amanullah} et~al.(2010){Amanullah}, {Lidman} et~al.}]{Union}
{Amanullah}, R., {Lidman}, C., {Rubin}, D., et~al., 2010, \apj, 716, 712,
  \eprint arXiv:{1004.1711}

\bibitem[{{Astier} et~al.(2006){Astier}, {Guy} et~al.}]{Astier06}
{Astier}, P., {Guy}, J., {Regnault}, N., et~al., 2006, \aap, 447, 31,
  \eprint{astro-ph/0510447}

\bibitem[{Bautista et~al.(2020)Bautista, Paviot et~al.}]{Bautista20}
Bautista, J.~E., Paviot, R., Vargas Magaña, M., et~al., 2020, Monthly Notices
  of the Royal Astronomical Society, 500, 1, 736–762, ISSN 1365-2966

\bibitem[{{Betoule} et~al.(2014){Betoule}, {Kessler} et~al.}]{Betoule14}
{Betoule}, M., {Kessler}, R., {Guy}, J., et~al., 2014, \aap, 568, A22, \eprint
  arXiv:{1401.4064}

\bibitem[{{Brout} et~al.(2021){Brout}, {Hinton} \& {Scolnic}}]{Binning}
{Brout}, D., {Hinton}, S.~R., {Scolnic}, D., 2021, \apjl, 912, 2, L26, \eprint
  arXiv:{2012.05900}

\bibitem[{{Brout} \& {Scolnic}(2021)}]{BS20}
{Brout}, D., {Scolnic}, D., 2021, \apj, 909, 1, 26, \eprint arXiv:{2004.10206}

\bibitem[{{Brout} et~al.(2022){Brout}, {Scolnic} et~al.}]{Brout22}
{Brout}, D., {Scolnic}, D., {Popovic}, B., et~al., 2022, arXiv e-prints,
  arXiv:2202.04077, \eprint arXiv:{2202.04077}

\bibitem[{Brout et~al.(2021)Brout, Taylor et~al.}]{fragilistic}
Brout, D., Taylor, G., Scolnic, D., et~al., 2021, The Pantheon+ Analysis:
  SuperCal-Fragilistic Cross Calibration, Retrained SALT2 Light Curve Model,
  and Calibration Systematic Uncertainty, \eprint arXiv:{2112.03864}

\bibitem[{{Campbell} et~al.(2013){Campbell}, {D'Andrea} et~al.}]{Campbell13}
{Campbell}, H., {D'Andrea}, C.~B., {Nichol}, R.~C., et~al., 2013, \apj, 763,
  88, \eprint arXiv:{1211.4480}

\bibitem[{Chabanier et~al.(2021)Chabanier, Etourneau et~al.}]{Chabanier21}
Chabanier, S., Etourneau, T., Goff, J.-M.~L., et~al., 2021, The Completed
  SDSS-IV extended Baryon Oscillation Spectroscopic Survey: The Damped
  Lyman-$\alpha$ systems Catalog, \eprint arXiv:{2107.09612}

\bibitem[{{Chambers} et~al.(2016){Chambers}, {Magnier} et~al.}]{Chambers16}
{Chambers}, K.~C., {Magnier}, E.~A., {Metcalfe}, N., et~al., 2016, ArXiv
  e-prints, \eprint arXiv:{1612.05560}

\bibitem[{{Childress} et~al.(2014){Childress}, {Wolf} \& {Zahid}}]{Childress14}
{Childress}, M.~J., {Wolf}, C., {Zahid}, H.~J., 2014, \mnras, 445, 1898,
  \eprint arXiv:{1409.2951}

\bibitem[{Collaboration(in prep)}]{DES5YR}
Collaboration, D. E.~S., in prep

\bibitem[{{Conley} et~al.(2011){Conley}, {Guy} et~al.}]{Conley11}
{Conley}, A., {Guy}, J., {Sullivan}, M., et~al., 2011, \apjs, 192, 1, \eprint
  arXiv:{1104.1443}

\bibitem[{{Doi} et~al.(2010){Doi}, {Tanaka} et~al.}]{Doi10}
{Doi}, M., {Tanaka}, M., {Fukugita}, M., et~al., 2010, \aj, 139, 1628, \eprint
  arXiv:{1002.3701}

\bibitem[{{Foley} et~al.(2018){Foley}, {Scolnic} et~al.}]{Foley18}
{Foley}, R.~J., {Scolnic}, D., {Rest}, A., et~al., 2018, \mnras, 475, 1, 193,
  \eprint arXiv:{1711.02474}

\bibitem[{{Frieman} et~al.(2008){Frieman}, {Bassett} et~al.}]{Frieman08}
{Frieman}, J.~A., {Bassett}, B., {Becker}, A., et~al., 2008, \aj, 135, 338,
  \eprint arXiv:{0708.2749}

\bibitem[{{Guillochon} et~al.(2018{\natexlab{a}}){Guillochon}, {Nicholl}
  et~al.}]{SNII_3}
{Guillochon}, J., {Nicholl}, M., {Villar}, V.~A., et~al., 2018{\natexlab{a}},
  \\apjs, 236, 6, \eprint arXiv:{1710.02145}

\bibitem[{{Guillochon} et~al.(2018{\natexlab{b}}){Guillochon}, {Nicholl}
  et~al.}]{SNIbc_3}
{Guillochon}, J., {Nicholl}, M., {Villar}, V.~A., et~al., 2018{\natexlab{b}},
  \\apjs, 236, 6, \eprint arXiv:{1710.02145}

\bibitem[{{Guy} et~al.(2010){Guy}, {Sullivan} et~al.}]{Guy10}
{Guy}, J., {Sullivan}, M., {Conley}, A., et~al., 2010, \aap, 523, A7, \eprint
  arXiv:{1010.4743}

\bibitem[{Hinton \& Brout(2020)}]{PIPPIN}
Hinton, S., Brout, D., 2020, Journal of Open Source Software, 5, 2122

\bibitem[{{Hlozek} et~al.(2012){Hlozek}, {Kunz} et~al.}]{Hlozek12}
{Hlozek}, R., {Kunz}, M., {Bassett}, B., et~al., 2012, \apj, 752, 79, \eprint
  arXiv:{1111.5328}

\bibitem[{{Holtzman} et~al.(2008){Holtzman}, {Marriner} et~al.}]{Holtzman08}
{Holtzman}, J.~A., {Marriner}, J., {Kessler}, R., et~al., 2008, \aj, 136, 2306,
  \eprint arXiv:{0908.4277}

\bibitem[{Hou et~al.(2020)Hou, Sánchez et~al.}]{Hou20}
Hou, J., Sánchez, A.~G., Ross, A.~J., et~al., 2020, Monthly Notices of the
  Royal Astronomical Society, 500, 1, 1201–1221, ISSN 1365-2966

\bibitem[{{Jha}(2017)}]{SNIax_1}
{Jha}, S.~W., 2017, {Type Iax Supernovae}, 375

\bibitem[{{Jones} et~al.(2018){Jones}, {Riess} et~al.}]{Jones18}
{Jones}, D.~O., {Riess}, A.~G., {Scolnic}, D.~M., et~al., 2018, \apj, 867, 108,
  \eprint arXiv:{1805.05911}

\bibitem[{Jones et~al.(2019)Jones, Scolnic et~al.}]{Jones19}
Jones, D.~O., Scolnic, D.~M., Foley, R.~J., et~al., 2019, The Astrophysical
  Journal, 881, 1, 19, ISSN 1538-4357

\bibitem[{Jönsson et~al.(2010)Jönsson, Sullivan et~al.}]{Jonsson10}
Jönsson, J., Sullivan, M., Hook, I., et~al., 2010, Monthly Notices of the
  Royal Astronomical Society, ISSN 1365-2966

\bibitem[{{Kelsey} et~al.(2020){Kelsey}, {Sullivan} et~al.}]{Kelsey20}
{Kelsey}, L., {Sullivan}, M., {Smith}, M., et~al., 2020, arXiv e-prints,
  arXiv:2008.12101, \eprint arXiv:{2008.12101}

\bibitem[{{Kelsey} et~al.(2022){Kelsey}, {Sullivan} et~al.}]{Kelsey22}
{Kelsey}, L., {Sullivan}, M., {Wiseman}, P., et~al., 2022, \mnras, \eprint
  arXiv:{2208.01357}

\bibitem[{{Kenworthy} et~al.(2021){Kenworthy}, {Jones} et~al.}]{Kenworthy21}
{Kenworthy}, W.~D., {Jones}, D.~O., {Dai}, M., et~al., 2021, \apj, 923, 2, 265,
  \eprint arXiv:{2104.07795}

\bibitem[{{Kessler} et~al.(2010{\natexlab{a}}){Kessler}, {Bassett}
  et~al.}]{SNII_1}
{Kessler}, R., {Bassett}, B., {Belov}, P., et~al., 2010{\natexlab{a}},
  \\Publications of the Astronomical Society of the Pacific, 122, 1415, \eprint
  arXiv:{1008.1024}

\bibitem[{{Kessler} et~al.(2010{\natexlab{b}}){Kessler}, {Bassett}
  et~al.}]{SNIbc_1}
{Kessler}, R., {Bassett}, B., {Belov}, P., et~al., 2010{\natexlab{b}},
  \\Publications of the Astronomical Society of the Pacific, 122, 1415, \eprint
  arXiv:{1008.1024}

\bibitem[{{Kessler} et~al.(2009{\natexlab{a}}){Kessler}, {Becker}
  et~al.}]{Kessler09}
{Kessler}, R., {Becker}, A.~C., {Cinabro}, D., et~al., 2009{\natexlab{a}},
  \apjs, 185, 32, \eprint arXiv:{0908.4274}

\bibitem[{{Kessler} et~al.(2009{\natexlab{b}}){Kessler}, {Bernstein}
  et~al.}]{SNANA}
{Kessler}, R., {Bernstein}, J.~P., {Cinabro}, D., et~al., 2009{\natexlab{b}},
  Publications of the Astronomical Society of the Pacific, 121, 1028, \eprint
  arXiv:{0908.4280}

\bibitem[{{Kessler} et~al.(2013){Kessler}, {Guy} et~al.}]{Kessler13}
{Kessler}, R., {Guy}, J., {Marriner}, J., et~al., 2013, \apj, 764, 48, \eprint
  arXiv:{1209.2482}

\bibitem[{{Kessler} et~al.(2019){Kessler}, {Narayan} et~al.}]{Kessler19}
{Kessler}, R., {Narayan}, G., {Avelino}, A., et~al., 2019, Publications of the
  Astronomical Society of the Pacific, 131, 1003, 094501, \eprint
  arXiv:{1903.11756}

\bibitem[{{Kessler} \& {Scolnic}(2017)}]{Kessler16}
{Kessler}, R., {Scolnic}, D., 2017, \apj, 836, 56, \eprint arXiv:{1610.04677}

\bibitem[{{Kessler} et~al.(in prep){Kessler}, {Vincenzi} \&
  {Armstrong}}]{Kessler23}
{Kessler}, R., {Vincenzi}, M., {Armstrong}, P., in prep

\bibitem[{{Kunz} et~al.(2007){Kunz}, {Bassett} \& {Hlozek}}]{Kunz07}
{Kunz}, M., {Bassett}, B.~A., {Hlozek}, R.~A., 2007, \prd, 75, 103508, \eprint
  arXiv:{astro-ph/0611004}

\bibitem[{{Marriner} et~al.(2011){Marriner}, {Bernstein} et~al.}]{Marriner11}
{Marriner}, J., {Bernstein}, J.~P., {Kessler}, R., et~al., 2011, \apj, 740, 72,
  \eprint arXiv:{1107.4631}

\bibitem[{{M{\"o}ller} \& {de Boissi{\`e}re}(2019)}]{Moller19}
{M{\"o}ller}, A., {de Boissi{\`e}re}, T., 2019, arXiv e-prints,
  arXiv:1901.06384, \eprint arXiv:{1901.06384}

\bibitem[{{Perlmutter} et~al.(1999){Perlmutter}, {Aldering}
  et~al.}]{Perlmutter99}
{Perlmutter}, S., {Aldering}, G., {Goldhaber}, G., et~al., 1999, \apj, 517,
  565, \eprint{astro-ph/9812133}

\bibitem[{{Peterson} et~al.(2021){Peterson}, {Kenworthy} et~al.}]{Peterson21}
{Peterson}, E.~R., {Kenworthy}, W.~D., {Scolnic}, D., et~al., 2021, arXiv
  e-prints, arXiv:2110.03487, \eprint arXiv:{2110.03487}

\bibitem[{{Pierel} et~al.(2018{\natexlab{a}}){Pierel}, {Rodney}
  et~al.}]{SNII_2}
{Pierel}, J.~D.~R., {Rodney}, S., {Avelino}, A., et~al., 2018{\natexlab{a}},
  \\Publications of the Astronomical Society of the Pacific, 130, 11, 114504,
  \eprint arXiv:{1808.02534}

\bibitem[{{Pierel} et~al.(2018{\natexlab{b}}){Pierel}, {Rodney}
  et~al.}]{SNIbc_2}
{Pierel}, J.~D.~R., {Rodney}, S., {Avelino}, A., et~al., 2018{\natexlab{b}},
  \\Publications of the Astronomical Society of the Pacific, 130, 11, 114504,
  \eprint arXiv:{1808.02534}

\bibitem[{{Planck Collaboration} et~al.(2020){Planck Collaboration}, {Aghanim}
  et~al.}]{Planck18}
{Planck Collaboration}, {Aghanim}, N., {Akrami}, Y., et~al., 2020, \aap, 641,
  A6, \eprint arXiv:{1807.06209}

\bibitem[{{Popovic} et~al.(2021{\natexlab{a}}){Popovic}, {Brout}, {Kessler} \&
  {Scolnic}}]{Popovic22}
{Popovic}, B., {Brout}, D., {Kessler}, R., {Scolnic}, D., 2021{\natexlab{a}},
  arXiv e-prints, arXiv:2112.04456, \eprint arXiv:{2112.04456}

\bibitem[{{Popovic} et~al.(2021{\natexlab{b}}){Popovic}, {Brout}, {Kessler},
  {Scolnic} \& {Lu}}]{Popovic21a}
{Popovic}, B., {Brout}, D., {Kessler}, R., {Scolnic}, D., {Lu}, L.,
  2021{\natexlab{b}}, \apj, 913, 1, 49, \eprint arXiv:{2102.01776}

\bibitem[{{Popovic} et~al.(2019){Popovic}, {Scolnic} \& {Kessler}}]{Popovic19}
{Popovic}, B., {Scolnic}, D., {Kessler}, R., 2019, arXiv e-prints,
  arXiv:1910.05228, \eprint arXiv:{1910.05228}

\bibitem[{{Qu} et~al.(2021){Qu}, {Sako}, {M{\"o}ller} \& {Doux}}]{Qu21}
{Qu}, H., {Sako}, M., {M{\"o}ller}, A., {Doux}, C., 2021, \aj, 162, 2, 67,
  \eprint arXiv:{2106.04370}

\bibitem[{{Riess} et~al.(1998){Riess}, {Filippenko} et~al.}]{Riess98}
{Riess}, A.~G., {Filippenko}, A.~V., {Challis}, P., et~al., 1998, \aj, 116,
  1009, \eprint{astro-ph/9805201}

\bibitem[{Ross et~al.(2015)Ross, Samushia, Howlett, Percival, Burden \&
  Manera}]{Ross15}
Ross, A.~J., Samushia, L., Howlett, C., Percival, W.~J., Burden, A., Manera,
  M., 2015, Monthly Notices of the Royal Astronomical Society, 449, 1, 835,
  ISSN 0035-8711,
  \eprint{https://academic.oup.com/mnras/article-pdf/449/1/835/13767551/stv154.pdf}

\bibitem[{{Sako} et~al.(2008){Sako}, {Bassett} et~al.}]{Sako08}
{Sako}, M., {Bassett}, B., {Becker}, A., et~al., 2008, \aj, 135, 348, \eprint
  arXiv:{0708.2750}

\bibitem[{{Sako} et~al.(2014){Sako}, {Bassett} et~al.}]{Sako16}
{Sako}, M., {Bassett}, B., {Becker}, A.~C., et~al., 2014, ArXiv e-prints,
  \eprint arXiv:{1401.3317}

\bibitem[{{Sako} et~al.(2018){Sako}, {Bassett} et~al.}]{Sako18}
{Sako}, M., {Bassett}, B., {Becker}, A.~C., et~al., 2018, Publications of the
  Astronomical Society of the Pacific, 130, 064002, \eprint arXiv:{1401.3317}

\bibitem[{{Sako} et~al.(2011){Sako}, {Bassett} et~al.}]{Sako11}
{Sako}, M., {Bassett}, B., {Connolly}, B., et~al., 2011, \apj, 738, 162,
  \eprint arXiv:{1107.5106}

\bibitem[{{S{\'a}nchez} et~al.(2021){S{\'a}nchez}, {Kessler}
  et~al.}]{Sanchez21}
{S{\'a}nchez}, B., {Kessler}, R., {Scolnic}, D., et~al., 2021, arXiv e-prints,
  arXiv:2111.06858, \eprint arXiv:{2111.06858}

\bibitem[{{Schlafly} \& {Finkbeiner}(2011)}]{Schlafly11}
{Schlafly}, E.~F., {Finkbeiner}, D.~P., 2011, \apj, 737, 103, \eprint
  arXiv:{1012.4804}

\bibitem[{{Scolnic} \& {Kessler}(2016)}]{Scolnic16}
{Scolnic}, D., {Kessler}, R., 2016, \apjl, 822, L35, \eprint arXiv:{1603.01559}

\bibitem[{{Scolnic} et~al.(2018){Scolnic}, {Kessler} et~al.}]{Scolnic18}
{Scolnic}, D., {Kessler}, R., {Brout}, D., et~al., 2018, \apjl, 852, L3,
  \eprint arXiv:{1710.05845}

\bibitem[{{Smith} et~al.(2020){Smith}, {Sullivan} et~al.}]{Smith20}
{Smith}, M., {Sullivan}, M., {Wiseman}, P., et~al., 2020, arXiv e-prints,
  arXiv:2001.11294, \eprint arXiv:{2001.11294}

\bibitem[{{Sullivan} et~al.(2010){Sullivan}, {Conley} et~al.}]{Sullivan10}
{Sullivan}, M., {Conley}, A., {Howell}, D.~A., et~al., 2010, \mnras, 406, 782,
  \eprint arXiv:{1003.5119}

\bibitem[{{Sullivan} et~al.(2011){Sullivan}, {Guy} et~al.}]{Sullivan11}
{Sullivan}, M., {Guy}, J., {Conley}, A., et~al., 2011, \apj, 737, 102, \eprint
  arXiv:{1104.1444}

\bibitem[{{Tripp}(1998)}]{Tripp98}
{Tripp}, R., 1998, \aap, 331, 815

\bibitem[{{Villar} et~al.(2017{\natexlab{a}}){Villar}, {Berger}, {Metzger} \&
  {Guillochon}}]{SNII_4}
{Villar}, V.~A., {Berger}, E., {Metzger}, B.~D., {Guillochon}, J.,
  2017{\natexlab{a}}, \\apj, 849, 70, \eprint arXiv:{1707.08132}

\bibitem[{{Villar} et~al.(2017{\natexlab{b}}){Villar}, {Berger}, {Metzger} \&
  {Guillochon}}]{SNIbc_4}
{Villar}, V.~A., {Berger}, E., {Metzger}, B.~D., {Guillochon}, J.,
  2017{\natexlab{b}}, \\apj, 849, 70, \eprint arXiv:{1707.08132}

\bibitem[{{Vincenzi} et~al.(2019){Vincenzi}, {Sullivan} et~al.}]{Vincenzi19}
{Vincenzi}, M., {Sullivan}, M., {Firth}, R.~E., et~al., 2019, \mnras, 489, 4,
  5802, \eprint arXiv:{1908.05228}

\bibitem[{{Vincenzi} et~al.(2021){Vincenzi}, {Sullivan} et~al.}]{Vincenzi21}
{Vincenzi}, M., {Sullivan}, M., {M{\"o}ller}, A., et~al., 2021, arXiv e-prints,
  arXiv:2111.10382, \eprint arXiv:{2111.10382}

\bibitem[{{Wiseman} et~al.(2022){Wiseman}, {Vincenzi} et~al.}]{Wiseman22}
{Wiseman}, P., {Vincenzi}, M., {Sullivan}, M., et~al., 2022, \mnras, 515, 3,
  4587, \eprint arXiv:{2207.05583}

\bibitem[{{Zuntz} et~al.(2015){Zuntz}, {Paterno} et~al.}]{zuntz}
{Zuntz}, J., {Paterno}, M., {Jennings}, E., et~al., 2015, Astronomy and
  Computing, 12, 45, \eprint arXiv:{1409.3409}

\end{thebibliography}

\end{document}